\documentclass[preprint]{aastex}

\usepackage{natbib}
\usepackage{color}
\newcommand{\beq}{\begin{equation}}
\newcommand{\beqa}{\begin{eqnarray}}
\newcommand{\eeq}{\end{equation}}
\newcommand{\eeqa}{\end{eqnarray}}

\newcommand{\simgt}{\lower.5ex\hbox{$\; \buildrel > \over \sim \;$}}
\newcommand{\simlt}{\lower.5ex\hbox{$\; \buildrel < \over \sim \;$}}

\newcommand{\bd}[1]{\mbox{\boldmath $#1$}}

\shorttitle{Systematic errors in weak-lensing}
\shortauthors{Shirasaki and Yoshida}

\begin{document}

\title{Statistical and Systematic Errors in Measurement of Weak-Lensing Minkowski Functionals: 
Application to Canada-France-Hawaii Lensing Survey
}

\author{Masato Shirasaki}
\affil{Department of Physics, University of Tokyo, Tokyo 113-0033, Japan}
\email{masato.shirasaki@utap.phys.s.u-tokyo.ac.jp}

\and 

\author{Naoki Yoshida}
\affil{
Department of Physics, University of Tokyo, Tokyo 113-0033, Japan\\
Kavli Institute for the Physics and Mathematics of the Universe (WPI),
University of Tokyo, Kashiwa, Chiba 277-8583, Japan
}

\begin{abstract}
The measurement of cosmic shear using weak gravitational lensing
is a challenging task that involves a number of complicated
procedures. We study in detail the systematic errors in  
the measurement of weak-lensing Minkowski Functionals (MFs).
Specifically, we focus on systematics associated
with galaxy shape measurements, photometric redshift errors,
and shear calibration correction.
We first generate mock weak-lensing catalogs 
that directly incorporate the actual observational characteristics 
of the Canada-France-Hawaii Lensing Survey (CFHTLenS).
We then perform a Fisher analysis 
using the large set of mock catalogs for various cosmological models.
We find that the statistical error associated with the 
observational effects degrades the cosmological parameter constraints 
by a factor of a few.
The Subaru Hyper Suprime-Cam (HSC) survey with a sky coverage 
of $\sim1400$ ${\rm deg}^2$ will constrain the dark energy 
equation of the state parameter with an error 
of $\Delta w_0 \sim 0.25$ by the lensing MFs alone, 
but biases induced by the systematics can be comparable to 
the $1\sigma$ error. 
We conclude that the lensing MFs are powerful statistics 
beyond the two-point statistics,
only if well-calibrated measurement of both
the redshifts and the shapes of source galaxies is performed.
Finally, we analyze the CFHTLenS data to explore 
the ability of the MFs to break degeneracies between 
a few cosmological parameters. 
Using a combined analysis of the MFs and the shear correlation function,
we derive the matter density $\Omega_{\rm m0} = 0.256\pm^{0.054}_{0.046}$.
\end{abstract}



\section{INTRODUCTION}

An array of recent observations such as
cosmic microwave background (CMB) anisotropies 
\citep[e.g.,][]{2011ApJS..192...18K,2013ApJS..208...19H} 
and large-scale structure 
\citep[e.g.,][]{2006PhRvD..74l3507T,2010MNRAS.404...60R}
established the standard $\Lambda$CDM model.
The energy content of the present-day universe is
dominated by dark energy and dark matter, and
the primordial density fluctuations, which seeded all rich 
structure we observe today,
were generated through an inflationary epoch in the very early universe. 
A few important questions still remain such as the 
nature of dark energy, the physical properties of dark matter, 
and the exact mechanism that generated
the primordial density fluctuations.

Gravitational lensing is a powerful method for studying matter 
distribution in the universe, from which one can extract  
information on the basic cosmological parameters.
The two-point correlation of galaxy ellipticities and the 
cross-correlation between galaxy positions
and the shear caused by the underlying matter 
are popular statistics for constraining the cosmological model.
These local statistics have been applied to real observational data
and have provided independent and comparable cosmological constraints 
to those obtained using other measurements such as galaxy clustering
\citep[e.g.,][]{2013MNRAS.430.2200K,2013MNRAS.432.1544M}.
Future weak-lensing surveys are aimed at measuring cosmic shear
over a wide area of more than 1000 ${\rm deg}^2$. 
Such observational programs include 
Subaru Hyper Suprime-Cam (HSC),\footnotemark[1]  
the Dark Energy Survey (DES),\footnotemark[2] 
and the Large Synoptic Survey Telescope (LSST).\footnotemark[3]
\footnotetext[1]{\rm{http://www.naoj.org/Projects/HSC/j\_index.html}} 
\footnotetext[2]{\rm{http://www.darkenergysurvey.org/}}
\footnotetext[3]{\rm{http://www.lsst.org/lsst/}}
The large set of cosmic shear data will enable us to improve 
the constraints on cosmological parameters which provides
important clues to the mysterious dark components.

Several groups have explored statistics that make the best use of 
cosmic shear for cosmological parameter constraints.
Minkowski Functionals (MFs) are among the most useful statistics to extract 
non-Gaussian information from a two-dimensional or three-dimensional field.
\citet{2001ApJ...552L..89M} and \citet{2001ApJ...551L...5S}
study $\Omega_{\rm m0}$ dependence of weak-lensing MFs.
More recently, \citet{2012PhRvD..85j3513K} show that the lensing 
MFs contain significant cosmological information, beyond the
power spectrum, whereas
\citet{2012ApJ...760...45S} show that weak-lensing MFs 
can be used to constrain
the statistical properties of the primordial density fluctuations.

The true applicability of morphological statistics on 
observational cosmic-shear data needs to be further explored in detail.
Previous studies on weak-lensing MFs often consider idealized 
conditions, assuming identical source redshift,
homogeneous angular distribution of sources
and/or perfectly calibrated shape measurement.
Many observational effects in real weak-lensing 
measurement exist, however, that can be sources of systematic errors,
for example, an imperfect shape measurement due to seeing and optical distortion, 
selection effects of source galaxies, uncertain redshift
distribution of the source galaxies due to photometric redshift error
\citep[e.g.,][]{2000A&amp;A...363..476B},
noise-rectification biases
\citep[e.g.,][]{2000ApJ537555K,2001A&A366717E,2003MNRAS.343..459H},
and complicated survey geometry due to mask regions \citep{2013ApJ...774..111S}.
Some of these effects on cosmic-shear power spectrum analysis 
have been already studied
\citep[e.g.,][]{2006MNRAS.366..101H,2011MNRAS.412...65H}.
It is timely and important to conduct 
a comprehensive study of observational effects on lensing MFs,
in order to fully exploit the data from 
upcoming wide-cosmology surveys.

Earlier in \citet{2013ApJ...774..111S},
we studied the effect of mask regions on the measurement of 
weak-lensing MFs. Geometrical constraints due to masks 
can be a major source of systematics because MFs 
are intrinsically morphological quantities.
We showed that sky-masking induces large non-Gaussianities,
which could compromise measurement of
the {\it true} non-Gaussianity associated with gravitational growth.
We thus argue that it is important to include directly
realistic observational effects 
in order to apply the lensing MFs to data from future cosmology
surveys. 

In the present paper, we further explore several 
observational effects using the real data set from 
the Canada-France-Hawaii Lensing Survey (CFHTLenS). 
We use a large set of simulations to study possible systematics
in detail one by one. We finally present a forecast for future surveys
such as Subaru HSC and LSST.
The rest of the present paper is organized as follows.
In Section~\ref{sec:WL}, we summarize the basics of lensing 
statistics of interest 
and how to estimate MFs from the observed shear field.
In Section~\ref{sec:data}, we describe the data used in this paper and 
the details of weak-lensing mock catalogs for morphological analysis 
are found in Section~\ref{sec:sim}.
In Section~\ref{sec:forecast}, we show the results of the impact of 
observational effects on lensing MFs. We perform a Fisher analysis to 
present a realistic forecast by using a large set of
weak-lensing mock catalogs.
In Section~\ref{sec:app}, we apply our statistical method 
to real data to quantify the power of
lensing MFs as a cosmology probe.
Concluding remarks and discussions are given in Section~\ref{sec:con}.

\section{WEAK-LENSING STATISTICS}\label{sec:WL}
We summarize the basics of gravitational lensing by large-scale structure.
Let us denote the observed position of a source object as $\bd{\theta}$ 
and the true position as $\bd{\beta}$.
Then the image distortion of a source object is characterized 
by the following two-dimensional (2D) matrix:
\beqa
A_{ij} = \frac{\partial \beta^{i}}{\partial \theta^{j}}
           \equiv \left(
\begin{array}{cc}
1-\kappa -\gamma_{1} & -\gamma_{2}  \\
-\gamma_{2} & 1-\kappa+\gamma_{1} \\
\end{array}
\right), \label{distortion_tensor}
\eeqa
where $\kappa$ is convergence and $\gamma = \gamma_1 + i \gamma_2$ is shear.
In a weak-lensing regime (i.e., $\kappa, \gamma \ll 1$), 
each component of $A_{ij}$ can be related to
the second derivative of the lensing potential 
$\Phi$ \citep{Bartelmann:1999yn,Munshi:2006fn}.
The lensing potential is calculated from the weighted integral 
of gravitational potential along a line of sight, and
the Poisson equation relates the gravitational potential field
to the matter over-density field $\delta$. 
Weak-lensing convergence field is then given by
\beqa
\kappa(\bd{\theta},\chi) &=& \int_{0}^{\chi_{H}}{\rm d}\chi \, W(\chi) \delta(r(\chi)\bd{\theta},\chi), 
\label{eq:kappa_delta} \\
W(\chi) &=& \frac{3}{2}\left(\frac{H_{0}}{c}\right)^2\Omega_{\rm m0}\frac{r(\chi)}{a(\chi)}
\int_{\chi}^{\chi_{H}}{\rm d}\chi^{\prime} p(\chi^{\prime})\frac{r(\chi^{\prime}-\chi)}{r(\chi^{\prime})},
\label{eq:weight_kappa}
\eeqa
where $\chi$ is the comoving coordinate, $r(\chi)$ is the angular diameter distance, 
and $p(\chi)$ represents the redshift distribution of sources.

\subsection{Two Point Statistics}
\label{sec:2pcf}
By using the Limber approximation \citep{Limber:1954zz,Kaiser:1991qi}
and Equation~(\ref{eq:kappa_delta}), 
one can calculate the convergence power spectrum as 
\beqa
P_{\kappa}(\ell) &=& \int_{0}^{\chi_s} {\rm d}\chi \frac{W(\chi)^2}{r(\chi)^2} 
P_{\delta}\left(k=\frac{\ell}{r(\chi)},z(\chi)\right)
\label{eq:kappa_power},
\eeqa
where $P_{\delta}(k)$ is the three-dimensional matter power spectrum.
The most direct measurement of weak-lensing two-point statistics is the
two-point shear correlation function (2PCF) $\xi_{\pm}$.
Theoretically, the 2PCFs are obtained by 
the Hankel transforms of a convergence power spectrum $P_{\kappa}$ as 
\beqa
\xi_{\pm}(\theta) = \frac{1}{2\pi}\int_{0}^{\infty}{\rm d}\ell \, \ell\, P_{\kappa}(\ell) \,J_{0,4}(\ell \theta),
\eeqa
where $J_{0}$ and $J_{4}$ are the first-kind Bessel functions of 
the order 0 and 4.

\citet{2002A&A...396....1S} 
show that the 2PCFs are estimated 
in an unbiased way by averaging over pairs of galaxies.
In practice, the estimator $\hat{\xi}_{\pm}$ is calculated by
\beqa
\hat{\xi}_{\pm}(\theta) = \frac{\sum_{ij}w_{i}w_{j}\left(\epsilon_{t}(\bd{\theta}_{i})\epsilon_{t}(\bd{\theta}_{j})
+\epsilon_{\times}(\bd{\theta}_{i})\epsilon_{\times}(\bd{\theta}_{j})\right)}{\sum_{ij}w_{i}w_{j}},
\eeqa
where $w_{i}$ is weight related to shape measurement and 
$\epsilon_{t,\times}(\bd{\theta}_{i})$ is the tangential and cross 
component of $i$th source galaxy's ellipticity.
The summation is taken over all galaxy pairs $(i,j)$ 
with angular separation 
$|\bd{\theta}_{i}-\bd{\theta}_{j}| \in [\theta-\Delta \theta, 
\theta+\Delta \theta]$.

\subsection{Minkowski Functionals}
\label{sec:MFs}
MFs are morphological statistics for some 
smoothed random fields characterized by a certain threshold.
In general, for a given $D$-dimensional smoothed field $\mathbb{S}^{D}$,  
one can calculate $D+1$ MFs $V_{i}$.
One can thus define 2+1 MFs 
on $\mathbb{S}^2$: $V_{0}, V_{1}$ and $V_{2}$.
The MFs are defined, for a threshold of $\nu$, as
\begin{eqnarray}
V_{0}(\nu) &\equiv& \frac{1}{4\pi}\int_{Q_{\nu}}\, {\rm d}S, \label{eq:V0_def} \\
V_{1}(\nu) &\equiv& \frac{1}{4\pi} \int_{\partial Q_{\nu}}\, \frac{1}{4} {\rm d}\ell , \label{eq:V1_def} \\
V_{2}(\nu) &\equiv& \frac{1}{4\pi} \int_{\partial Q_{\nu}}\, \frac{1}{2\pi}K{\rm d}\ell , \label{eq:V2_def}
\end{eqnarray}
where $Q_{\nu}$ and $\partial Q_{\nu}$ represent the excursion set 
and its boundaries for a smoothed field $u(\bd{\theta})$.
They are given by
\beqa
Q_{\nu} = \{\bd{\theta}\, |\, u(\bd{\theta}) > \nu\}, \\
\partial Q_{\nu} =\{ \bd{\theta} \, |\, u(\bd{\theta}) = \nu \}.
\eeqa
For a given threshold, $V_{0}$, $V_{1}$ and $V_{2}$ describe 
the fraction of sky area,
the total boundary length of contours, 
and the integral of the geodesic curvature $K$ along the contours,
respectively.

\subsection{Measuring Lensing MFs from Cosmic Shear Data}
\label{sec:est_MF}
We summarize how to estimate lensing MFs from observed cosmic shear.
Let us first define the weak-lensing mass map, i.e.,
the smoothed lensing convergence field:
\beqa
{\cal K} (\bd{\theta}) = \int {\rm d}^2 \phi \ \kappa(\bd{\theta}-\bd{\phi}) U(\bd{\phi}), \label{eq:ksm_u}
\eeqa
where $U$ is the filter function to be specified below.
We can calculate the same quantity by smoothing the shear field $\gamma$ as
\beqa
{\cal K} (\bd{\theta}) = \int {\rm d}^2 \phi \ \gamma_{t}(\bd{\phi}:\bd{\theta}) Q_{t}(\bd{\phi}), \label{eq:ksm}
\eeqa
where $\gamma_{t}$ is the tangential component of the shear at position $\bd{\phi}$ relative to the
point $\bd{\theta}$.
The filter function for the shear field $Q_{t}$ is related to $U$ by
\beqa
Q_{t}(\theta) = \int_{0}^{\theta} {\rm d}\theta^{\prime} \ \theta^{\prime} U(\theta^{\prime}) - U(\theta).
\label{eq:U_Q_fil}
\eeqa
We consider $Q_{t}$ to be defined with a finite extent.
In this case, one finds 
\beqa
U(\theta) = 2\int_{\theta}^{\theta_{o}} {\rm d}\theta^{\prime} \ \frac{Q_{t}(\theta^{\prime})}{\theta^{\prime}} - Q_{t}(\theta),
\eeqa
where $\theta_{o}$ is the outer boundary of the filter function.

In the following, we consider the truncated Gaussian filter (for $U$) as
\beqa
U(\theta) &=& \frac{1}{\pi \theta_{G}^{2}} \exp \left( -\frac{\theta^2}{\theta_{G}^2} \right)
-\frac{1}{\pi \theta_{o}^2}\left( 1-\exp \left(-\frac{\theta_{o}^2}{\theta_{G}^2} \right) \right), \\
Q_{t}(\theta) &=& \frac{1}{\pi \theta^{2}}\left[ 1-\left(1+\frac{\theta^2}{\theta_{G}^2}\right)\exp\left(-\frac{\theta^2}{\theta_{G}^2}\right)\right],
\label{eq:filter_gamma}
\eeqa
for $\theta \leq \theta_{o}$ and $U = Q_{t} = 0$ elsewhere.
Throughout the present paper, we adopt $\theta_{G} = 1$ arcmin 
and $\theta_{o} = 15$ arcmin.
Note that this choice of $\theta_{G}$ is considered to be
an optimal smoothing scale for the detection of massive galaxy clusters
using weak-lensing for $z_{\rm source}$ = 1.0 \citep{2004MNRAS.350..893H}.

It is important to use appropriately the weight associated 
with shape measurement when making smoothed convergence maps.
In practice, we estimate ${\cal K}$ by 
generalizing Equation~(\ref{eq:ksm}):
\beq
{\cal K}(\bd{\theta}_{i}) = \frac{\sum_{j}Q_{t}(\bd{\phi}_{j})w_{j}
\epsilon_{t}(\bd{\phi}_{j}:\bd{\theta}_{i})}{\sum_{j}Q_{t}(\bd{\phi}_{j})w_{j}},
\label{eq:ksm_prac}
\eeq 
where the summation in Equation~(\ref{eq:ksm_prac}) is taken over 
all the source galaxies that are located within $\theta_{o}$ from $i$th pixel.
The weak-lensing convergence field ${\cal K}$ is then computed 
from the galaxy ellipticity data 
on regular grids with a grid spacing of 0.15 arcmin.
In making the convergence map, we discard the pixels
when the denominator in Equation~(\ref{eq:ksm_prac}) is equal to zero.
The boundaries are defined by masking a pixel 
if the number of sources within $\theta_{o}$ 
from the pixel is less than $5\sqrt{15\pi \theta_{o}^2}$.
This critical value effectively sets the signal-to-noise ratio of the 
number of sources inside a circle with radius of $\theta_{o}$ to be less than 5, 
on the assumption that the distribution of sources is approximated by a Poisson 
distribution. We repeat the above procedure for all the pixels.
Note that the details of the procedure do not affect the final 
results significantly
as long as we impose the same conditions to on all of the pixels,
because our analysis is based on the comparison
of two maps that have the same configuration of source positions.

We follow \citet{2012JCAP...01..048L}
to calculate the MFs from pixelated ${\cal K}$ maps.
We convert a weak-lensing field ${\cal K}$ to $x = ({\cal K} - \langle {\cal K} \rangle)/\sigma_{0}$
where $\sigma_{0}$ is the standard deviation of ${\cal K}$.
We set $\Delta x = 0.2$ from $x=-5$ to $x = 5$ for binning the threshold value.

\section{DATA}
\label{sec:data}
We use the data from
the Canada-France-Hawaii Telescope Lensing Survey \citep[CFHTLenS;][]{2012MNRAS.427..146H}.
CFHTLenS is a 154 ${\rm deg}^2$ multi-color optical survey in 
the five optical bands
$u^{*}, g^{\prime}, r^{\prime}, i^{\prime}, \ {\rm and} \ z^{\prime}$.
CFHTLenS is optimized for weak-lensing analysis with a full 
multi-color depth of 
$i^{\prime}_{AB} = 24.7$ with optimal sub-arcsecond seeing conditions.
The survey consists of four separated regions called W1, W2, W3 and W4, 
with an area of $\sim$ 72, 30, 50 and 25 ${\rm deg}^2$, respectively.

The CFHTLenS survey analysis mainly consists of the following processes:  
weak-lensing data processing with THELI
\citep{2013MNRAS.433.2545E}, 
shear measurement with the $lens$fit
\citep{2013MNRAS.429.2858M}, 
and photometric redshift measurement
\citep{2012MNRAS.421.2355H}.
A detailed systematic error study of the shear measurements 
in combination with the photometric redshifts 
is presented in \citet{2012MNRAS.427..146H}.
The additional error analyses of the photometric redshift 
measurements are presented in 
\citet{2013MNRAS.431.1547B}.

The ellipticities of the source galaxies in the data have been 
calculated using the $lens$fit algorithm.
The $lens$fit performs a Bayesian model fitting to the 
imaging data by varying a galaxy's ellipticity
and size, and by marginalizing over the centroid position.
It takes into account a forward convolution process 
expressed by convolving the galaxy model
with the point-spread function (PSF) to estimate the posterior 
probability of the model given the data.
For each galaxy, the ellipticity $\epsilon$ is estimated
as the mean likelihood of the model posterior
probability after marginalizing over galaxy size, 
centroid position, and bulge fraction.
An inverse variance weight $w$ is given by the variance 
of the ellipticity likelihood surface
and the variance of the ellipticity distribution of the galaxy population
(see \cite{2013MNRAS.429.2858M} for further details).

The photometric redshifts $z_{p}$ are estimated by the {\tt BPZ} code 
\cite[][Bayesian Photometric Redshift Estimation]{2000ApJ...536..571B}.
\citet{2013MNRAS.431.1547B} shows that 
the true redshift distribution is well 
described by the sum of the probability distribution functions (PDFs) 
estimated from {\tt BPZ}.
The galaxy-galaxy-lensing redshift scaling analysis 
of \citet{2012MNRAS.427..146H}
confirms that contamination is unimportant for galaxies selected 
at $0.2 < z_{p} < 1.3$.
In this redshift range, the weighted median redshift is $\sim0.7$ and 
the effective weighted number density $n_{\rm eff}$ is 11 per 
${\rm arcmin}^2$.
We have used the source galaxies with $0.2 < z_{p} < 1.3$
to make the smoothed lensing mass map analyzed in \ref{sec:est_MF}.

\begin{figure}[!t]
\begin{center}
       \includegraphics[clip, width=0.7\columnwidth]{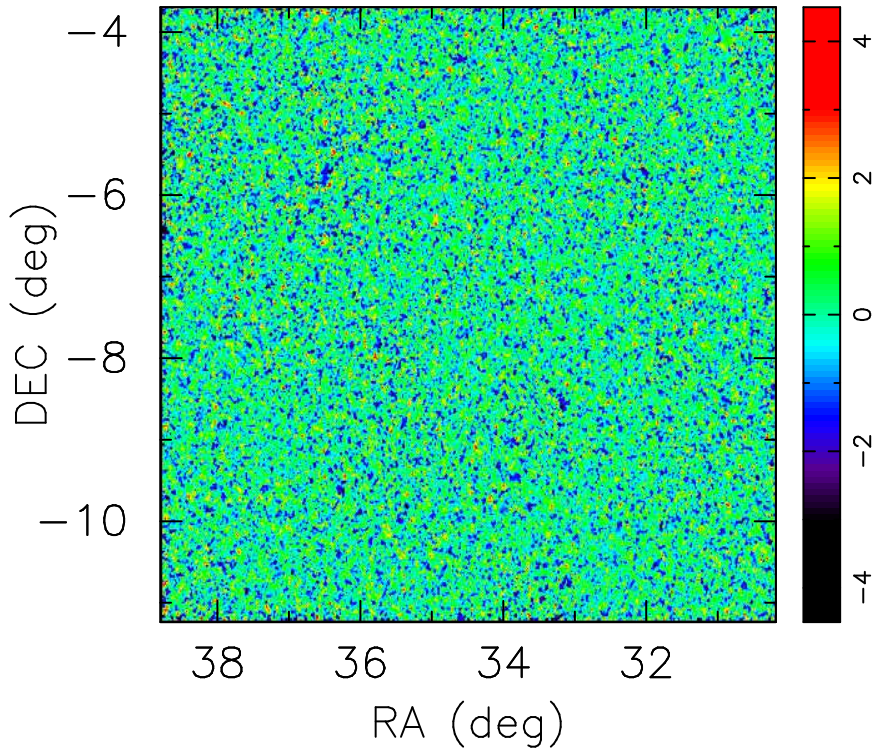}
    \caption{
    	Reconstructed convergence field ${\cal K}$
        in the CFHTLenS W1 field.
	The ${\cal K}$ map is calculated from the ellipticities 
        of 2570270 source galaxies.
	The color-scale bar shows the normalized value 
        $({\cal K} - \langle {\cal K} \rangle)/\sigma_{0}$.
     } 
     \label{fig:W1map}
    \end{center}    
\end{figure}

The effective survey area is an important quantity for our study.
\citet{2012MNRAS.427..146H} perform systematic tests in order to mark 
and remove data with significant residual systematics.
The fraction of data flagged by their procedure amounts to 25 \% of the total CFHTLenS; 
this is indeed significant. Our previous work \citep{2013ApJ...774..111S} 
shows that the cosmological information content in the lensing MFs 
is largely determined by the effective survey area. 
More importantly, however, complicated geometries of the masked 
regions induce non-Gaussianities that contaminate the lensing MFs.
We have decided to use all the available data of CFHTLenS 
to make a wide {\it and} continuous map.
We expect the systematics associated with the PSF to be
relatively small compared to the masking effect on morphological 
statistics (see, e.g., \cite{2012MNRAS.427..146H, 2013MNRAS.433.3373V}). 
To calculate the 2PCF, we use the {\it clean} sample 
of \citet{2012MNRAS.427..146H}.
We show the obtained mass map in the CFHTLenS W1 field in Figure \ref{fig:W1map}.

\section{SIMULATIONS}
\label{sec:sim}
\subsection{Ray-Tracing Simulation}
\label{subsec:rt}

We first run a number of cosmological $N$-body simulations to generate
a three-dimensional matter density field. We use the parallel Tree-Particle Mesh 
code {\tt Gadget2}
\citep{2005MNRAS.364.1105S}. The simulations are run with $512^3$ dark matter 
particles in a volume of $480$ or $960 \ h^{-1}$Mpc on a side. 
We generate the initial conditions using a parallel code 
developed by \citet{2009PASJ...61..321N} and
\citet{2011A&A...527A..87V}, which employs the 
second-order Lagrangian perturbation theory 
\cite[e.g.,][]{2006MNRAS.373..369C}.
The initial redshift is set to 
 $z_{\rm init}=50$, where we compute the linear matter transfer function using
 {\tt CAMB} \citep{Lewis:1999bs}.
Our fiducial cosmology adopts the following parameters:
matter density $\Omega_{\rm m0}=0.279$, dark energy density $\Omega_{\Lambda 0}=0.721$, 
the amplitude of curvature fluctuations 
$A_{s}=2.41\times 10^{-9}$ at the pivot scale $k=0.002{\rm Mpc}^{-1}$,
the parameter of the equation of state of dark energy $w_{0} = -1$,
Hubble parameter $h=0.700$ and 
the scalar spectral index $n_s=0.972$.
These parameters are consistent with 
the WMAP nine-year results \citep{2013ApJS..208...19H}.
To investigate the degeneracy of the cosmological parameters, 
we also run the same set of simulations but with slightly 
different $\Omega_{\rm m0}$, $w_0$ and $A_{s}$.
We summarize the simulation parameters in Table \ref{tab:nbody}.

For ray-tracing simulations of gravitational lensing, 
we generate light-cone outputs using multiple simulation boxes
in the following manner. Our small- and large-volume simulations are 
placed to cover the past light-cone of a hypothetical observer 
with an angular extent $10^{\circ}\times 10^{\circ}$, from 
$z=0$ to $3$, similarly to the methods in  
\citet{2000ApJ...537....1W},
\citet{2001MNRAS.327..169H},
and
\citet{2009ApJ...701..945S}.
Details of the configuration are found in the last reference.
The angular grid size of our maps is 
$10^{\circ}/4096\sim 0.15$ arcmin.
We use outputs from independent
realizations when generating the light-cone outputs.
We also randomly shift the simulation boxes
in order to avoid the same structure appearing multiple times
along a line-of-sight.
In total, we generate 40 independent shear maps 
from four $N$-body simulations for each cosmological model.

\begin{table}[!t]
\begin{center}
\begin{tabular}{|c|c|c|c|c|c|c|c|}
\tableline
& $\Omega_{\rm m0}$ & $w_{0}$ & $A_{s}\times 10^9$ & $\sigma_{8} $&\# of $N$-body sims & \# of maps\\ \tableline
Fiducial & 0.279 & -1.0 & 2.41 & 0.823 & 4 & 40\\ \tableline
High $\Omega_{\rm m0}$ & 0.304 & -1.0 & 2.41 & 0.878 & 4 & 40\\ \tableline
Low $\Omega_{\rm m0}$ & 0.254 & -1.0 & 2.41 & 0.763 & 4 & 40\\ \tableline
High $w_0$ & 0.279 & -0.8 & 2.41 & 0.768 & 4 & 40\\ \tableline
Low $w_0$ & 0.279 & -1.2 & 2.41 & 0.862 & 4 & 40\\ \tableline
High $A_s$ & 0.279 & -1.0 & 2.51 & 0.840 & 4 & 40\\ \tableline
Low $A_s$ & 0.279 & -1.0 & 2.31 & 0.806 & 4 & 40\\ \tableline
\end{tabular} 
\caption{
Parameters for our $N$-body simulations.
We also show the resulting $\sigma_{8}$.
For each parameter set, we run 4 $N$-body 
realizations and generate 40 weak-lensing shear maps.
}
\label{tab:nbody}
\end{center}
\end{table}
\subsection{Mock Weak-Lensing Catalogs}
\label{subsec:mock}

Our purpose is to study observational effects on 
weak-lensing morphological statistics.
To this end, we generate realistic mock weak-lensing catalogs 
by combining ray-tracing simulations and the CFHTLenS data
\citep{2013MNRAS.433.3373V}.
The main advantage of our mock catalogs is that 
the observed positions on the sky of the source galaxies 
are directly used.
Because of this, we can keep all the characteristics 
of the survey geometry the same as in CFHTLenS.

We locate the source galaxies in the pixel unit of our lensing map 
and then calculate the reduced shear signal $g=\gamma/(1-\kappa)$ 
at the galaxy positions.
Ray-tracing is done up to the redshift of the galaxy
as described in Section~\ref{subsec:rt}.
In this step, a galaxy's redshift is set to 
be at the peak of the posterior PDF obtained from {\tt BPZ}.
This could cause systematic effects on 
morphological statistics originating from 
the inaccuracy of the photometric redshift estimation.
We discuss the impact of the redshift distribution 
of sources on lensing morphological statistics later in 
Section 5.

We next consider the intrinsic ellipticity that is 
known to be a major error source in cosmic shear measurement.
For each galaxy, we randomize the orientation of the observed ellipticity,
while keeping its amplitude.
The randomized ellipticity is then assigned as 
the intrinsic ellipticity $\epsilon_{\rm int}$ at each galaxy's position.
\citet{1997A&A...318..687S} show that  
the observed source ellipticity sheared by weak-lensing effect 
can be expressed as a function of $\epsilon_{\rm int}$ and the 
reduced shear signal $g$.
The final ``observed" ellipticity is 
\beqa
{\bd \epsilon}_{\rm mock} = \frac{{\bd \epsilon}_{\rm int}+{\bd g}}{1+{\bd g}^{*}{\bd \epsilon}_{\rm int}},
\eeqa
where ${\bd \epsilon}_{\rm mock}$ is represented as a complex ellipticity.

Finally, we incorporate calibration correction in the shear measurement.
We assign the weight associated with the shape measurement of $lens$fit 
and the shear calibration correction following
\citet{2012MNRAS.427..146H}.
The two factors determine the potential additive shear bias ${\bd c}$ 
and multiplicative bias $m$.
We then apply the shear calibration correction to ${\bd \epsilon}_{\rm mock}$ by using 
bias factors $m$ and ${\bd c}$ as
\beqa
{\bd \epsilon}_{\rm mock} \rightarrow (1+m){\bd \epsilon}_{\rm mock}+{\bd c}.
\eeqa
In this step, we assume that there is 
no correlation between ${\bd \epsilon}$ and $m, \bd{c}$.
We have explicitly calculated 
the correlation between ${\bd \epsilon}$ and $m, \bd{c}$
at the source galaxy positions using the CFHTLenS data set,
and found that there is indeed no significant correlation 
between the quantities.

Through the above procedures, we have successfully 
included the following observational 
effects in our analysis:
(1) non-linear relation between the 
observed ellipticities and cosmic shear, 
(2) non-Gaussian distribution of the intrinsic
ellipticities, 
(3) the masked survey area of CFHTLenS and 
the inhomogeneous angular distribution of the source galaxies,
(4) imperfect shape measurements
and (5) the redshift distribution of the source galaxies.
We note that all or many of these effects are 
often ignored in previous works 
on lensing morphological statistics.

\section{REALISTIC FORECAST}
\label{sec:forecast}
\subsection{Fisher Analysis}
\label{subsec:FA}
We perform a Fisher analysis to produce a forecast for parameter 
constraints on $\Omega_{\rm m0}$, $A_s$, and $w_0$
with future weak-lensing surveys.

\begin{figure}[!t]
\begin{center}
       \includegraphics[clip, width=0.32\columnwidth]{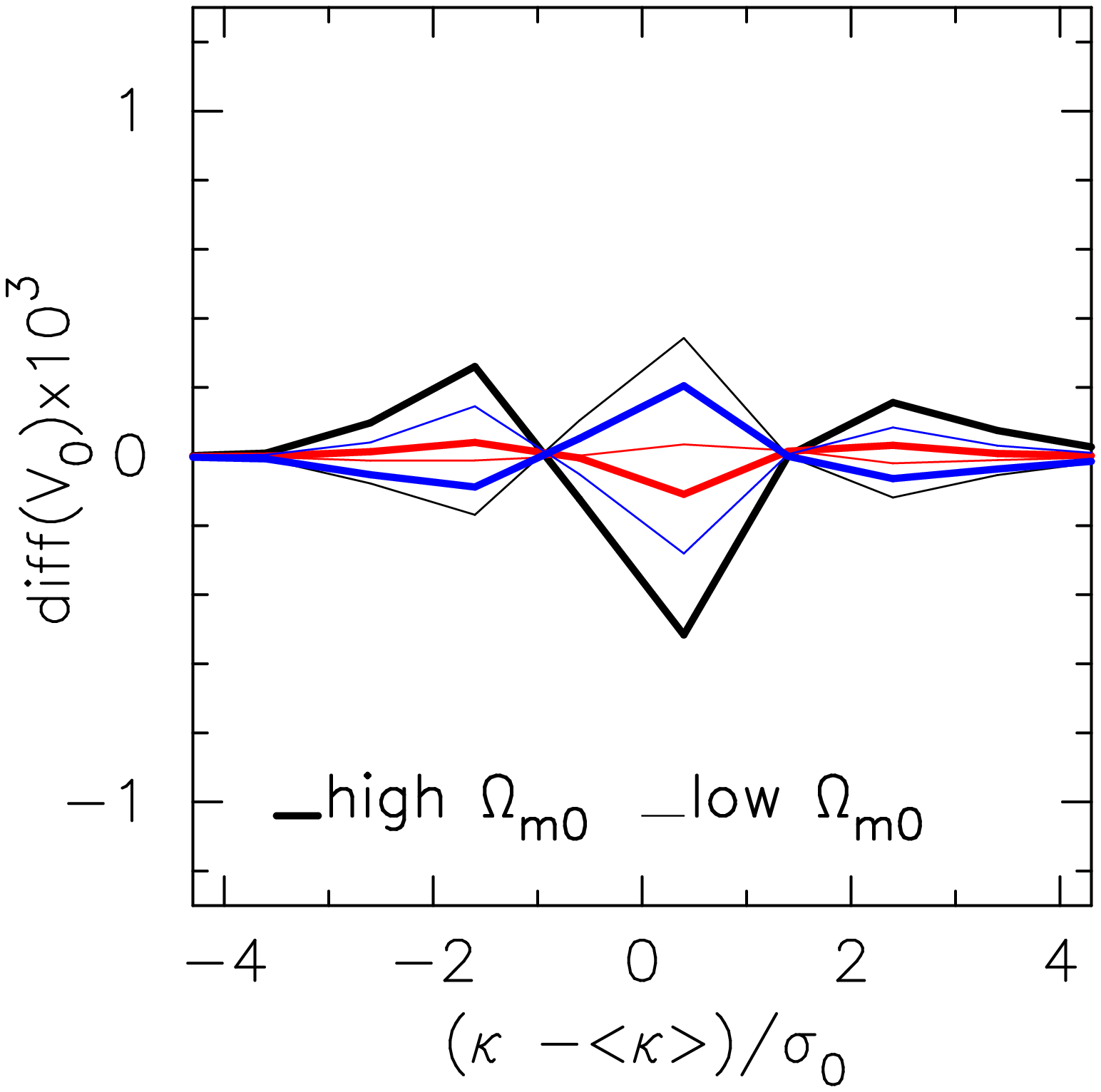}
       \includegraphics[clip, width=0.32\columnwidth]{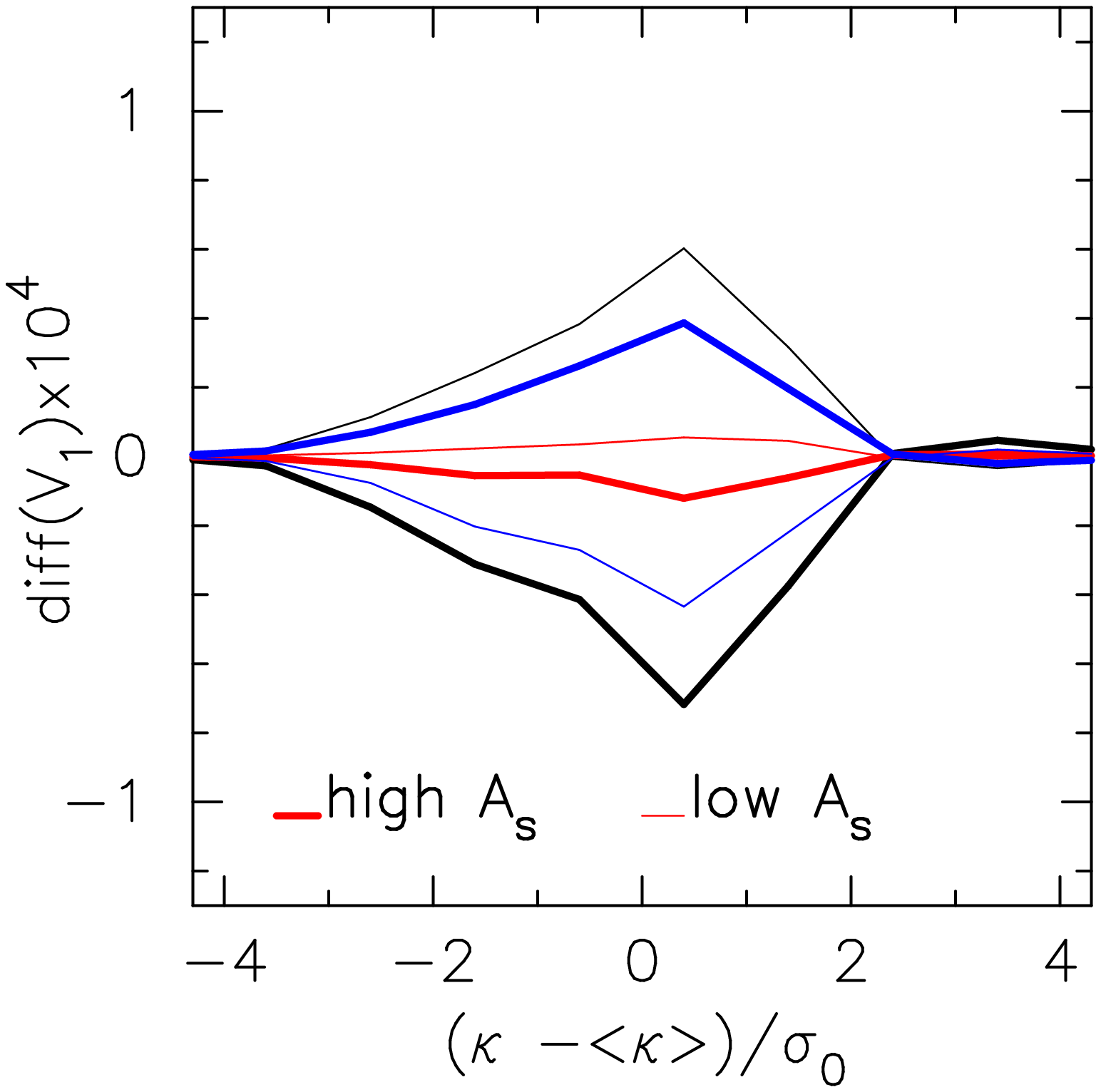}
       \includegraphics[clip, width=0.32\columnwidth]{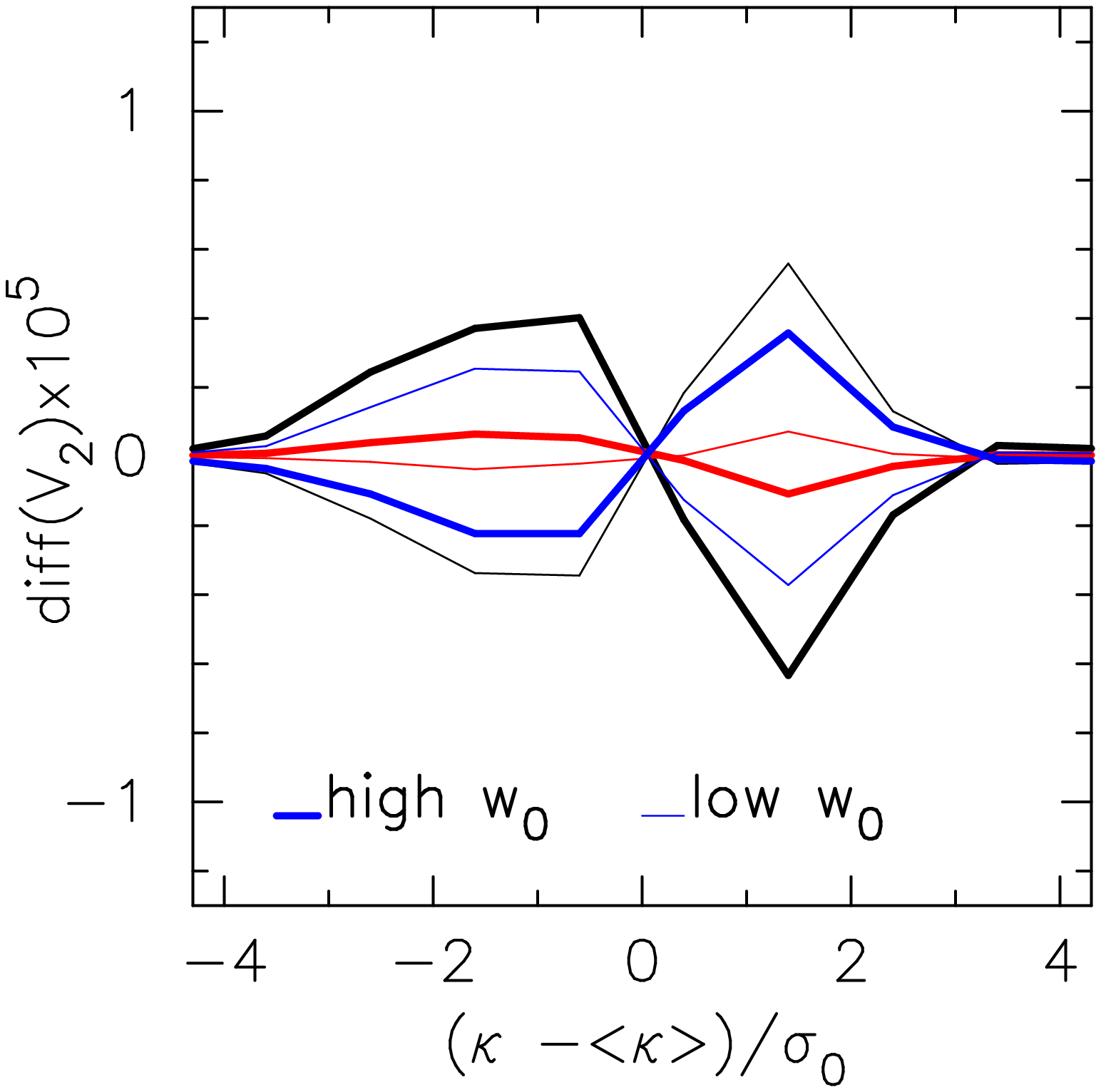}
    \caption{
	Variation of the lensing MFs for different cosmological parameters.
	We plot the differences of $V_0, V_1$, and $V_2$.
        with respect to those of 
        our fiducial cosmology. 
        In all of the panels, the thick (thin) black line corresponds to the case of 
        cosmological model 
	with higher (lower) $\Omega_{\rm m0}$.
	The thick (thin) red one shows the result of the cosmological model 
        with higher (lower) $A_s$,
	and the thick (thin) blue one is for the model with higher (lower) $w_0$.
        For reference, the typical statistical errors 
        of $V_{0}$,$V_{1}$, and $V_{2}$ 
	at $({\cal K} - \langle {\cal K} \rangle)/\sigma_{0}=0$ are 
	$\sim10^{-3}$, $10^{-4}$, and $10^{-5}$, respectively,
        for CFHTLenS.
     } 
     \label{fig:diffMFs}
    \end{center}
\end{figure}

For a multivariate Gaussian likelihood, the Fisher matrix $F_{ij}$ 
can be written as
\beqa
F_{ij} = \frac{1}{2} {\rm Tr} \left[ A_{i} A_{j} + C^{-1} M_{ij} \right], \label{eq:Fij}
\eeqa
where $A_{i} = C^{-1} \partial C/\partial p_{i}$, 
$M_{ij} = 2 \left(\partial \mu/\partial p_{i} \right)\left(\partial \mu/\partial p_{j} \right)$, 
$C$ is the data covariance matrix, 
$\mu$ represents the assumed model, 
and $\bd{p} = (\Omega_{\rm m0}, A_s, w_0)$ are the main parameters.
In the present study, we consider only the second term in Equation~(\ref{eq:Fij}).
Because $C$ is expected to scale proportionally inverse to the survey area, 
the second term will be dominant 
for a large area survey (see, e.g., \citet{2009A&A...502..721E}).
We calculate the model template by averaging the MFs over 
40 convergence maps with appropriate noises 
for each CFHTLenS field.
Figure \ref{fig:diffMFs} shows the cosmological dependence 
of our model MFs thus calculated. We see clearly the behavior
of the MFs as a function of $\bd{p}$.

We calculate the model 2PCFs using
the fitting formula of non-linear matter power spectrum 
of \citet{2012ApJ...761..152T},
on the assumption that the source redshift distribution 
is well approximated 
by the sum of the posterior PDF with $0.2 < z_p < 1.3$ given in
\citet{2013MNRAS.430.2200K}.

To calculate the matrix $M_{ij}$, we approximate the first derivatives of the 2PCF and MFs 
with respect to cosmological parameter $p_{i}$ as
\beqa
\frac{\partial \mu}{\partial p_{i}} = \frac{\mu(p^{(0)}_{i}+\Delta p_{i}) - \mu(p^{(0)}_{i} - \Delta p_{i})}{2\Delta p_{i}}, \label{eq:dev_MFs}
\eeqa
where $\bd{p}^{(0)}=(0.279, 2.41 \times 10^{-9}, -1.0)$ gives our fiducial model parameters 
and we set $\Delta \bd{p}=(0.025, 0.1 \times 10^{-9},0.2)$.

We construct the data vector ${\bd D}$ from a set of binned MFs and 2PCFs, 
\beqa
{D_{i}} = \{V_{0}(x_{1}),...,V_{0}(x_{10}),V_{1}(x_{1}),...,V_{1}(x_{10}),
V_{2}(x_{1}),...,V_{2}(x_{10}), \nonumber \\
\xi_{+}(\theta_{1}),...,\xi_{+}(\theta_{10}),
\xi_{-}(\theta_{1}),...,\xi_{-}(\theta_{10})\}, \label{eq:Dvec}
\eeqa
where $x_{i}=({\cal K}_{i} - \langle {\cal K} \rangle)/\sigma_{0}$ 
is the binned normalized lensing field.
For the Fisher analysis, we use 10 bins in the range of $x_{i} = [-3,3]$
\footnote{
In principle, one can use regions with $x>3$ as well.
Such regions usually correspond to the positions of massive dark 
matter halos, which are thought to be sensitive to cosmological 
parameters. On the other hand, such regions are extremely rare, and 
thus the first derivatives in Equation~(\ref{eq:dev_MFs}) 
are not evaluated accurately even with our large number of ${\cal K}$ maps.
We thus do not use high ${\cal K}$ regions with $x>3$ in our analysis.
}.
In this range of $x$, Equation~(\ref{eq:dev_MFs}) gives smooth 
estimates for $M_{ij}$.
For the 2PCFs, we use 10 bins logarithmically spaced 
in the range of $\theta_{i} = [0.9, 300]$ arcmin.
A data vector has 50 elements, 
$3\times10$ MFs and $2\times10$ 2PCFs, in total.

We therefore need a $50 \times 50$ data covariance matrix 
for the Fisher analysis.
To this end, we first use 1000 shear maps made 
by \citet{2009ApJ...701..945S}.
The maps have almost the same design as our simulations, 
but are generated for slightly different cosmological parameters 
(consistent with WMAP three-years results \citep{2007ApJS..170..377S}).
The actual parameter differences are small,
and also the dependence of the covariance matrix
on cosmological parameters is expected to be weak.
Each map covers a $5^{\circ} \times 5^{\circ}$ sky 
and the source 
redshift is set to be $z_{\rm source} = 1$.
We model the intrinsic ellipticities by adding random ellipticities drawn 
from a 2D Gaussian to the simulated shear data. 
We set the rms of the intrinsic ellipticities to be 0.38 
and the number of source galaxies is set to be 10 ${\rm arcmin}^{-2}$.
These are reasonable choices for our study here.
In making the smoothed lensing map from the simulation outputs, 
we set the weight related to shape measurement to be unity.
From the 1000 shear maps with appropriate noises, 
we can estimate the variances of the 2PCFs and MFs.
We also estimate the statistical errors 
by randomly rotating the observed orientation of the ellipticities.
Using these randomized catalogs, 
the data covariance matrices in each CFHTLenS field can be estimated
by the sum of sampling variance and the statistical error as
\beqa
C_{\rm each} = C_{\rm cosmic} \left(\frac{A_{\rm each} \ {\rm deg}^2}{25 \ {\rm deg}^2}\right)^{-1} + C_{\rm stat}, \label{eq:Cov}
\eeqa
where $C_{\rm cosmic}$ is sampling variance, 
$C_{\rm stat}$ represents the statistical error in each CFHTLenS field,
and $A_{\rm each}$ corresponds to the effective survey area.
In the following, we assume that the four 
CFHTLenS fields are independent of each other statistically.
The total inverse covariance matrix for the whole CFHTLenS data
is the sum of $C_{\rm each}^{-1}$ over the W1, W2, W3 and W4 fields.
We forecast for future lensing surveys by simply scaling 
the data covariances by the survey area, assuming that
the statistical error is identical to that in CFHTLenS.
When calculating the inverse covariance, 
we include a debiasing correction, the so-called 
Anderson-Hartlap factor $\alpha=(n_{\rm real}-n_{\rm bin}-2)/(n_{\rm real}-1)$ 
with $n_{\rm rea}=1000$ being the number of realization of simulation sets and
$n_{\rm bin}=50$ being the number of total bins in our data vector
\citep{2007A&A...464..399H}.

We expect that Equation~(\ref{eq:Cov}) provides a good approximation 
to the full covariance, but the accuracy needs to be addressed here. 
In the case of shear correlation functions, 
the covariance matrix consists of three components:
a sampling variance, the statistical noise, 
and a third term coupling the two \citep{2002A&A...396....1S}. 
Because the MFs do not have the additivity of,
e.g., $V_{i}(\nu_{1}+\nu_{2}) = V_{i}(\nu_{1})+V_{i}(\nu_{2})$,
it is not clear if the MF covariance can be expressed similarly 
as the sum of the three contributions.
We thus resort to estimating the MF covariance in a direct manner by
using the large set of mock catalogs generated by the procedure
shown in Section~\ref{subsec:mock}.
Note that, in the procedure, we perform two randomization processes
for a fixed cosmological model.
One is to generate multiple realizations of the large scale structure
(by $N$-body simulations) and the other is to randomize 
intrinsic ellipticities of the source galaxies.
We perform each process separately, technically by fixing 
a random seed of the other process, to evaluate each term 
in Equation~(\ref{eq:Cov}). 
In principle, one can perform both of the processes simultaneously
and derive the full covariance.
However, this would require a huge number of mock catalogs.
In Appendix B, we have done a simple but explicit check to validate 
that Equation~(\ref{eq:Cov}) indeed provides a reasonably good approximation.
The details of our test and the result are shown there. 

\subsection{Forecast for Upcoming Survey}
\label{subsec:forecast}
We present a forecast for upcoming surveys
such HSC and LSST.
We first derive constraints on the cosmological 
parameters for a 154 ${\rm deg}^2$ area survey, 
for which we have the full covariance matrix obtained 
in the previous sections.
We then consider two wide surveys with an area coverage 
of 1400 ${\rm deg}^2$ (HSC)
and 20000 ${\rm deg}^2$ (LSST).
We simply scale the covariance matrix by a factor 
of $154/1400$ or $154/20000$ for them.

\begin{figure}[!t]
\begin{center}
       \includegraphics[clip, width=0.5\columnwidth]{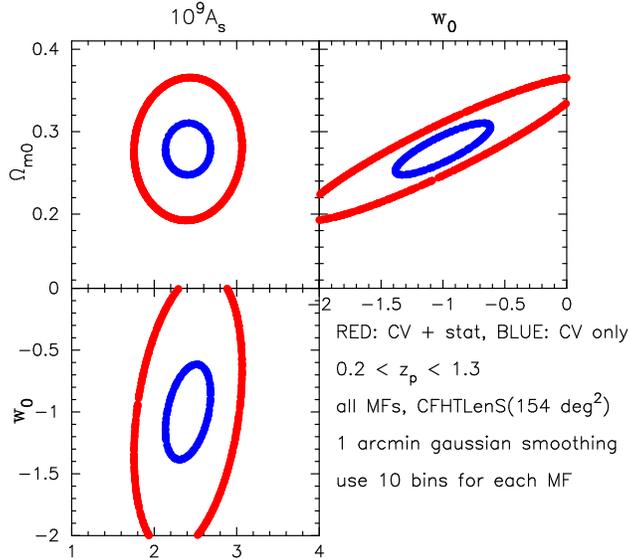}
    \caption{
    	Impact of statistical errors on the cosmological parameter estimation.
	We plot the 1$\sigma$ cosmological constraints by the lensing 
        MFs in the CFHTLenS case.
	The red error circle presents forecast that
	includes sampling variance and the statistical error associated 
        with the observational effects.
	The blue circle is obtained with only sampling variance included.
     } 
    \label{fig:impact_Cstat}
    \end{center}
\end{figure}

Let us begin with quantifying the statistical error 
associated with the real data.
We have performed a Fisher analysis including the sampling variance 
and the statistical error. 
When we include the statistical error,
the cosmological constraints are degraded by a factor of $\sim 2$
for the CHFTLens survey.
In Figure \ref{fig:impact_Cstat},
the red error circle corresponds to the 1$\sigma$ cosmological 
constraints including the sampling variance 
and the statistical error, while the blue one is obtained from the 
Fisher analysis without the statistical error.

\begin{figure}[!t]
\begin{center}
      \includegraphics[clip, width=0.5\columnwidth]{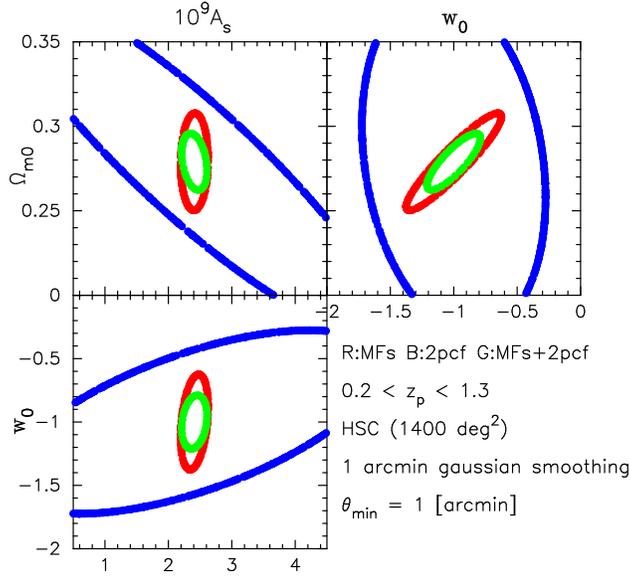}
    \caption{
    	Forecast for cosmological parameter constants 
        by lensing statistics 
	for the Subaru Hyper Suprime Cam survey (1400 ${\rm deg}^2$).
	The blue error circle represents the constraints from the 2PCFs, 
        whereas the red one is obtained from the MFs.
	Combining the two statistics can improve cosmological 
        constraints as indicated by the green circle.
	The data covariances for this plot are estimated from 1000 
        ray-tracing simulations 
	and 1000 randomized catalogs based on the CFHTLenS data.
     } 
    \label{fig:FA_2pcf+MFs}
    \end{center}
\end{figure}

We are now able to 
present a forecast for future lensing surveys covering larger sky areas
on the assumption that the data covariance is the same as that of CFHTLenS.
Figure \ref{fig:FA_2pcf+MFs} shows the derived parameter constraints.
The blue error circles show the 1$\sigma$ constraints from the shear 2PCFs 
whereas the red circles are obtained from the lensing MFs.
It is promising that, with Subaru HSC,  we can constrain the dark
energy equation of state $w_{0}$ with an 
error of $\Delta w_0 \sim 0.25$ by the lensing MFs alone.
Table \ref{tab:sig_forecast}  summarizes 
the expected constraints by future surveys.
Combining the 2PCFs and the MFs 
can improve the constraints by a factor of $\sim 2$ 
by breaking the degeneracy between the three parameters.
It should be noted that this conclusion might seem 
slightly different from that of \citet{2012PhRvD..85j3513K},
who argue that adding the power spectrum does not effectively improve 
the constraints when all three MFs are already used.
Our result suggests that combining the 2PCFs and the MFs 
improves cosmological parameter constraints appreciably.
A precise account for the difference is not given by
our study only, but there are many factors that can affect the
parameter constraint. First of all,
we characterize the amplitude of the matter power spectrum 
by the amplitude of curvature perturbations $A_{s}$ at
the cosmological recombination epoch
whereas \citet{2012PhRvD..85j3513K} adopt
$\sigma_{8}$ at the present epoch as a parameter. 
The latter is the so-called derived parameter and has an 
internal degeneracy with $\Omega_{\rm m}$ and $w_0$. 
Our result suggests that including the 2PCFs in the analysis 
can better constrain $A_{s}$,
which in turn yields tighter constraints on the other parameters.
Furthermore, our analysis includes observational effects 
such as survey mask regions and the source distribution 
directly. Altogether, these differences make it difficult
to compare our results with those of previous works that
mostly adopt idealized configurations.

\begin{table}[!t]
\begin{center}
\begin{tabular}{|c|c|c|c|}
\tableline
& $\Omega_{\rm m0}$ & $A_{s}\times 10^9$ & $w_{0}$ \\ \tableline
MFs only (1400 ${\rm deg}^2$) & 0.0190 &  0.143 & 0.248 \\ \tableline
MFs only (20000 ${\rm deg}^2$) & 0.00503 & 0.0380 & 0.0658 \\ \tableline
MFs + 2PCFs (1400 ${\rm deg}^2$) & 0.0110 & 0.132 & 0.139\\ \tableline
MFs + 2PCFs (20000 ${\rm deg}^2$) & 0.00293 & 0.0351 & 0.0369 \\ \tableline
\end{tabular} 
\caption{
  The 1$\sigma$ constraint on $\Omega_{\rm m0}$, $A_{s}$, and $w_{0}$,
  when marginalized 
  over the other two parameters.
  We consider two surveys with a survey area of 1400 ${\rm deg}^2$ (HSC)
  and 20000 ${\rm deg}^2$ (LSST).
  The analysis includes the sampling variance and the statistical error
  (see text). 
  \label{tab:sig_forecast}
}
\end{center}
\end{table}

\subsection{Possible Systematics}
\label{subsec:sys}
In this section, we examine the effect of known systematics on 
measurement of the MFs.
We follow \citet{2006MNRAS.366..101H} to estimate
the bias in the cosmological parameter due to 
some possible systematics 
\beqa
\delta p_{\alpha} = F^{-1}_{\alpha \beta}\sum_{i,j} 
C^{-1}_{ij}(D^{\rm test}_{i}-D^{\rm fid}_{i})
\frac{\partial D^{\rm fid}_{j}}{\partial p_{\beta}}
\label{eq:bias_CP}
\eeqa
where $\delta p_{\alpha}$ is the bias in the $\alpha$th cosmological parameter, 
$F_{\alpha \beta}$ is a Fisher matrix, ${\bd D}$ is the data vector and ${\bd C}$ 
is the data covariance.
The data vector ${\bd D}^{\rm fid}$ represents the theoretical template for our fiducial model.
and ${\bd D}^{\rm test}$ is the test data vector that includes a known systematics effect.
In this section, we use the data vector ${\bd D}$ consisting of the lensing MFs only.
For ${\bd D}^{\rm fid}$, we use the average MFs over 40 mock catalogs from our fiducial
cosmological model described in \ref{subsec:mock}.
The mock samples serve as reference, for which we have assumed that
(1) the source galaxy redshift is well approximated by the peak of the posterior 
PDF of photometric redshift,
and (2) the observed shear is perfectly calibrated by a functional form shown in 
\citet{2012MNRAS.427..146H}.
We test these assumptions and quantify the net effects
in a direct manner by generating and using another set of the mock catalogs 
for ${\bd D}^{\rm test}$ using the same $N$-body realizations as for our fiducial case.

\begin{figure}[!t]
\begin{center}
      \includegraphics[clip, width=0.4\columnwidth]{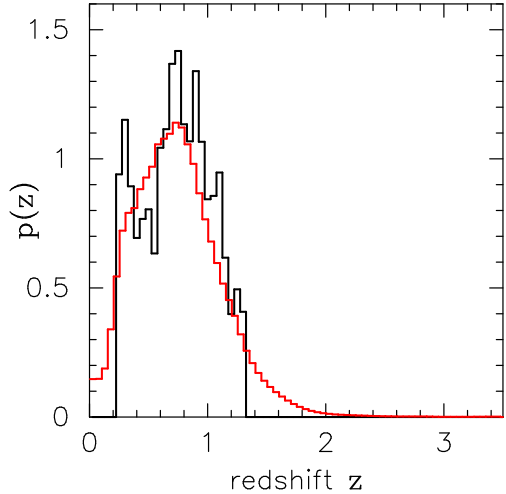}
    \caption{
	Redshift distribution function of sources $p(z)$.
	The red histogram shows the sum of the posterior PDF over galaxies with $0.2 < z_{p} < 1.3$.
	The black one is calculated from the peak value of the posterior PDF, i.e., the best-fit photometric redshift.
	The mean redshift is 0.69 for the black histogram and 0.748 for the red one.
     } 
    \label{fig:zdist}
    \end{center}
\end{figure}

\subsubsection*{Redshift Distribution}

It is important to quantify the effect of the source redshift 
distribution and of the error in photometric redshifts 
on the lensing MFs, or indeed on any lensing statistics.
We perform ray-tracing simulations by shooting 
rays to the farthest lens plane at $z=3$, 
weighting the lensing kernel using 
a redshift distribution function of the sources.
Specifically, we follow the same manner in as 
\citet{2009ApJ...701..945S} to 
simulate the weak-lensing effect
but the lensing kernel is slightly different from their simulation 
because of the wider source redshift distribution.
In \citet{2009ApJ...701..945S}, they assume $z_{p} = 1$ 
so that the lensing kernel can be calculated by the simple expression, 
i.e., $r(\chi_{s}-\chi_{l})r(\chi_{l})/r(\chi_{s})$,
where $\chi_{s}$ and $\chi_{l}$ are the comoving distance of sources and 
of the lens, respectively.
In the present study, we consider source redshift {\it distribution} $p(\chi)$.
Then the lensing kernel for the lensing objects at $\chi_{l}$ should be
replaced with
$\int_{\chi_{l}}^{\chi_{H}}{\rm d}\chi_{s}p(\chi_s)r(\chi_{s}-\chi_{l})r(\chi_{l})/r(\chi_{s})$.
The source positions on the sky and all the other characteristics
are kept the same as in our original mock catalogs,
which themselves are derived from CHFTLenS.
For the redshift distribution, 
we adopt the sum of the posterior PDF of photometric 
redshift for the galaxies with $0.2 < z_{p} < 1.3$.
Figure \ref{fig:zdist} compares 
the integrated redshift distribution with
the histograms of the source redshifts.
The latter is used in our fiducial simulations.
The test data vector ${\bd D}^{\rm test}$ is calculated by 
averaging the MFs over the new 40 catalogs with the posterior weight
described above.

The main difference caused by the different redshift distributions is 
the amplitude of the standard deviation of ${\cal K}$.
We see in  Figure \ref{fig:diffMFs_sys} that 
the net difference is as large as those found for cosmological 
models differing by $\Delta w_0 = 0.2$; this can obviously be 
a significant source of error in cosmological 
parameter constraints with upcoming future surveys.
We estimate the resulting bias in the derived $w_0$ 
by using Equation~(\ref{eq:bias_CP}).
The uncertainties in the photometric redshifts can indeed induce a
$\Delta w_0 \sim 0.1$ bias in $w_{0}$. 
The exact values are summarized in Table \ref{tab:cp_bias}.

We have also studied the effect of source 
redshift {\it clustering} on the lensing MFs.
The results are presented in Appendix A.
Briefly, the source redshift clustering is found to be a minor effect, 
but we note that it could cause non-negligible bias in ``precision" cosmology with, 
for example, the LSST lensing survey.

\subsubsection*{Shear Calibration Correction}

We next study the effect of shear calibration correction.
We consider the standard correction that describes the calibration 
as $\epsilon = (1+m){\bd \epsilon}_{\rm mock}+{\bd c}$ with
a multiplicative component $m$ and an additive component $c$.
The former is calibrated by analyzing simulated images whereas
the latter is calibrated empirically using the actual data. 
An ideal case would be one with $m=c=0$, which might possibly be
realized if a perfect calibration is done.
We compare the lensing maps with and without the calibration 
factors $m$ and $c$ in order to quantify the importance of the shear
calibration. 
We simply reanalyze the fiducial mock catalogs by 
setting $m=c=0$ for all of the source galaxies.
The resulting 40 mock catalogs are used to obtain
the data vector ${\bd D}^{\rm test}$ for this study.

\begin{figure}[!t]
\begin{center}
       \includegraphics[clip, width=0.32\columnwidth]{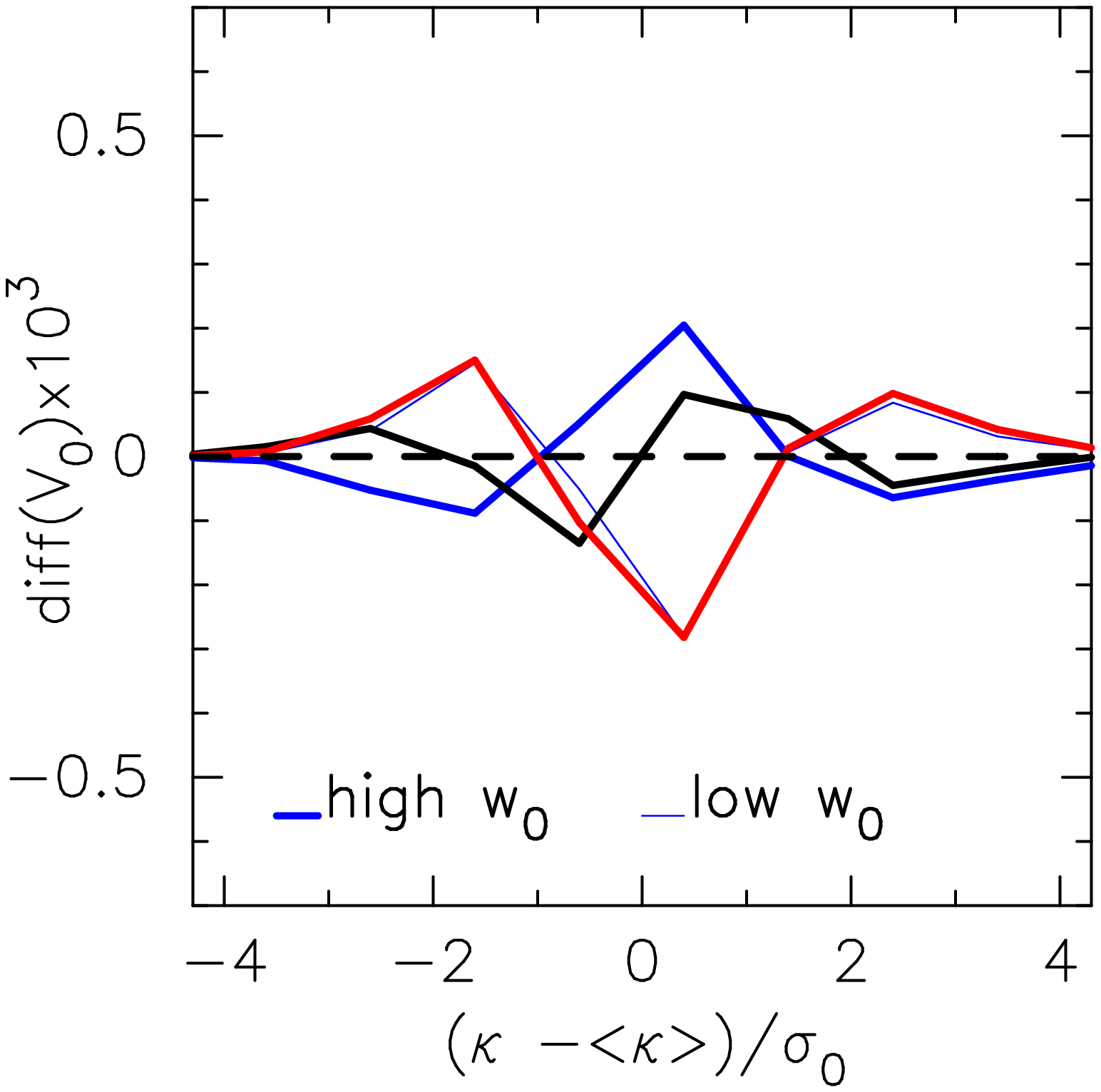}
       \includegraphics[clip, width=0.32\columnwidth]{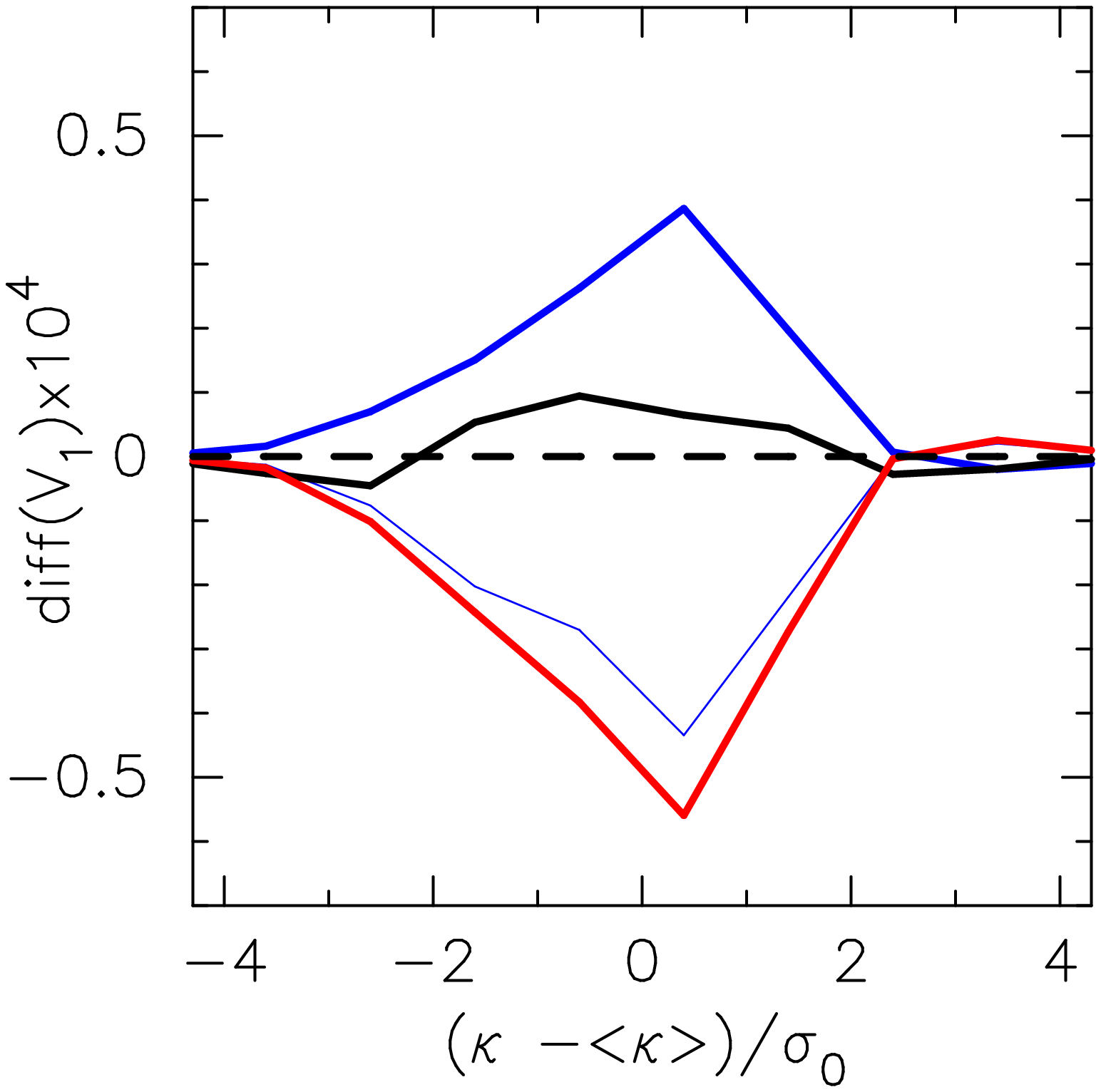}
       \includegraphics[clip, width=0.32\columnwidth]{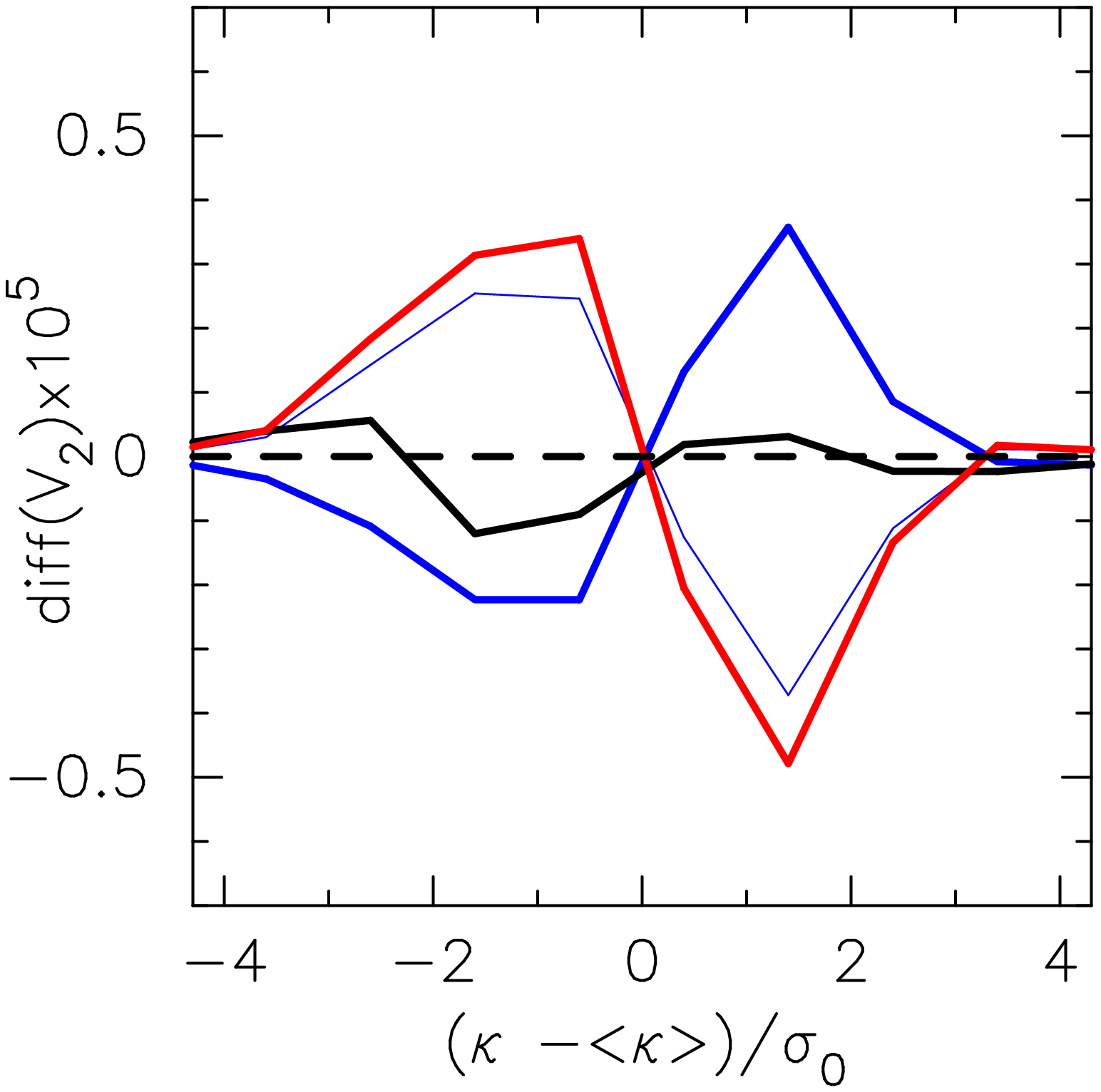}
    \caption{
	Impact of possible systematics on lensing MFs.
	We plot the differences of average MFs over 40 catalogs 
        between our fiducial cosmology and another one 
	that includes a given systematic.
	We generate a new set of mock catalogs in order to include systematics in our analysis.
	The red line shows the difference owing to source redshift distribution
	whereas the black one represents the effect of shear calibration 
        correction on the lensing MFs.
	For comparison, the thick (thin) blue line corresponds to the case 
        of a cosmological model 
	with higher (lower) $w_0$.
     } 
    \label{fig:diffMFs_sys}
    \end{center}
\end{figure}

We find that the additive calibration 
induces negligible effect 
but that the multiplicative calibration affects the 
lensing MFs appreciably.
In the case of CFHTLenS, the multiplicative calibration 
results in a $\sim6$ \% correction with
$\langle 1+m \rangle \simeq 0.94$.
Note that $m$ is a function of 
both the galaxy signal-to-noise ratio and the size.
Thus the calibration differs from position to position
and introduces effectively additional non-Gaussianities
to the ${\cal K}$ map.
Figure \ref{fig:diffMFs_sys} shows that the non-Gaussianities
actually cause biases in the lensing MFs.
The biases cannot simply be described by the difference 
of the standard deviation of ${\cal K}$, 
i.e., by the normalization
of the lensing MFs.
The resulting bias in the cosmological parameter estimate is close
to the $1\sigma$ level for an HSC-like survey
as shown in Table \ref{tab:cp_bias}.
Our study presented here suggests that 
the multiplicative correction needs to be included in $\it model$ 
predictions of the MFs for producing robust forecasts 
for upcoming surveys.
We emphasize that our theoretical templates are based 
on mock catalogs that directly include
the multiplicative correction
obtained from the real observational data.


\begin{table}[!t]
\begin{center}
\begin{tabular}{|c|c|c|c|}
\tableline
& $\Omega_{\rm m0}$ & $A_{s}\times 10^9$ & $w_{0}$ \\ \tableline
Redshift distribution & 0.00707 &  -0.0254 & -0.122 \\ \tableline
Calibration correction & -0.0224 & 0.110 & -0.234 \\ \tableline
\end{tabular} 
\caption{
  The bias of cosmological parameter estimation due to possible systematics.
  \label{tab:cp_bias}
}
\end{center}
\end{table}

\section{APPLICATION to CFHTLenS}
\label{sec:app}
We apply all the methods developed and examined
in the previous sections to the CFHTLenS data.
We have already shown in Section~\ref{subsec:forecast} 
that the statistical error in CFHTLenS degrades
the constraints on cosmological parameters if we
use only the lensing MFs.
It would be ideal to utilize other cosmological
probes to put tighter constraints.
We will use the CMB data from WMAP.

The likelihood analysis in this section includes the
systematics studied in Section~\ref{subsec:sys},
and so our result could be ``correctly'' biased.
Still, the lensing MFs are a powerful
cosmological probe, as we shall show in the following.


\subsection{Data Sets}
We use multiple data sets.
As a probe of large-scale structure, we use data from 
the nine-year WMAP data release
\citep{2013ApJS..208...20B,2013ApJS..208...19H}.
We use the output of Monte Carlo Markov Chains (MCMC) derived 
from the likelihood analysis with
the CMB temperature and polarization power- and cross-spectrum 
in the WMAP9 data.
Note that the MCMC we use here does not include other external data sets, 
such as small-scale CMB measurements, galaxy redshift surveys,
and Hubble constant.
We calculate the likelihood in our parameter space 
$\bd{p} = (\Omega_{\rm m0}, A_s, w_0)$
after marginalizing over the following three parameters: 
the reionization optical depth $\tau$, 
scalar spectral index $n_s$ and Hubble parameter $H_{0}$.

As a probe of matter distribution at low redshifts, 
we use the lensing MFs and the 2PCF calculated from the CFHTLenS data.
We construct the data vector and covariances 
in the same manner as in Section~\ref{subsec:FA},
but we assume no covariances between the MFs and the 2PCF.
To validate the assumption, we have actually calculated 
the parameter constraints 
by the Fisher analysis with/without covariances 
between the MFs and the 2PCF.
We have found that the approximation does not affect 
the final results significantly for the current data set.
The error in $\Omega_{\rm m0}$ increases only by $\Delta \Omega_{\rm m0} = 5 \times 10^{-4}$.

We sample the posterior of the cosmological parameters 
from the lensing 2PCF data set using the Population Monte Carlo (PMC) 
using the publicly available code 
${\tt COSMO\verb|_|PMC}$ \citep{2011arXiv1101.0950K}.
Details of the PMC are found in \citet{2009PhRvD..80b3507W}.
The method incorporates the cosmological dependence of the shear covariance 
in the manner described in \citet{2009A&A...502..721E}.
The same model parameters are adopted as in \citet{2013MNRAS.430.2200K},
with the smallest and largest angular bins being 0.9 and 300 arcmin.
We consider the following set of cosmological parameters:
$\bd{p}=({\Omega_{\rm m0}, \Omega_{\rm b0}, \sigma_{8}, H_{0}, n_s, w_0})$,
where $\Omega_{\rm b0}$ is the baryon density 
and $\sigma_{8}$ normalizes the matter power spectrum.
To compare the result derived from CMB and that from the lensing MFs, 
we calculate the value of $A_{s}$ at each sample point in parameter space
by using the following relation:
\beqa
A_{s} &=& A_{s, \rm fid}\left(\frac{\sigma_{8}}{\sigma_{8, \rm fid}}\right)^2
\frac{SD_{+}^2|_{\rm fid}}{SD_{+}^2}, \\
S &=& \int_{0}^{\infty} \frac{{\rm d}^3 k}{(2\pi)^3}k^{n_s}T(k)^2|W_8(k)|^2,
\eeqa
where $D_{+}$ is the linear growth factor of matter density,
$T(k)$ is the transfer function,
and $W_{8}(k)$ is the top-hat function with scale of 8 Mpc/h in the Fourier space. 
For the fiducial parameter set, we use the same parameters as the WMAP9 best-fit values.

In our PMC run, we perform 30 iterations to find a suitable 
importance function compared to the posterior.
Also 100,000 sample points are generated for each iteration.
To obtain a large sample set, we combine the PMC samples 
with the five highest value of perplexity $p$,
which is the conventional diagnostic that indicates 
the quality and effectiveness of the sampling.
Our PMC run achieves $p>0.7$ for the final samples;
this criterion is the same as that adopted in the analysis in
\citet{2013MNRAS.430.2200K}.

In the next sections, we study the following three cases:
(1) likelihood analysis with the lensing MFs alone,
(2) combined analysis with the lensing MFs and the 2PCF,
and (3) combined analysis with the lensing MFs and CMB anisotropies.
In the last analysis, we treat the lensing MFs data and the CMB data 
as being independent of each other.

\subsection{Likelihood Analysis of Lensing MFs}
In our maximum likelihood analysis,
we assume that the data vector ${\bd D}$ is well approximated 
by the multivariate Gaussian distribution with covariance ${\bd C}$.
This assumption is reasonable for the case of joint analysis 
of the CMB and lensing power spectrum
\citep{2010PhRvL.105y1301S}.
In this case, the $\chi^2$ statistics (log-likelihood) is given by
\beqa
\chi^2 = (D_{i}-\mu_{i}({\bd p}))C^{-1}(D_{j}-\mu_{j}({\bd p}))
\eeqa
where ${\bd \mu}({\bd p})$ is the theoretical prediction 
as a function of cosmological parameters.
The theoretical prediction is computed in a three-dimensional parameter space.
In sampling the likelihood function, we consider the 
limited parameter region as follows:
$\Omega_{\rm m0} \in [0,1]$, $A_{s}\times 10^9 \in [0.1,8.0]$ and $w_{0} \in [-6.5, 0.5]$.
The sampling number in each parameter is set to 100.

To estimate the MFs components in ${\bd \mu}$, 
we assume that the lensing MFs depend linearly
on the cosmological parameters with
the first derivatives calculated from Equation~(\ref{eq:dev_MFs}).
We consider two components of the contribution of the data covariance 
in our likelihood analysis; 
one is the statistical error and sampling variance, 
which are estimated as in Section~\ref{subsec:FA} while
the other originates from the possible systematics 
as studied in Section~\ref{subsec:sys}.
We denote the latter contribution as ${\bd C}^{\rm sys}$.
We estimate ${\bd C}^{\rm sys}$ in a simple and direct manner 
using the differences of the MFs,
as shown in Figure \ref{fig:diffMFs_sys}:
\beqa
C^{\rm sys}_{ij} = \left[(D^{\rm zdist}_{i}-D^{\rm fid}_{i})^2 
+ (D^{\rm scc}_{i}-D^{\rm fid}_{i})^2\right]\delta^{2D}_{ij},
\eeqa
where ${\bd D}^{\rm fid}$ is the template MFs for our fiducial mock catalogs,
${\bd D}^{\rm zdist}$ is the average MFs over 40 catalogs reflecting the different source redshift
distribution as shown in Figure \ref{fig:zdist},
and ${\bd D}^{\rm scc}$ is estimated by averaging the 
MFs over 40 catalogs without shear calibration correction.
The total covariance is the sum of the above two contributions.

\subsection{Breaking Degeneracies}
\begin{figure}[!t]
\begin{center}
      \includegraphics[clip, width=0.40\columnwidth]{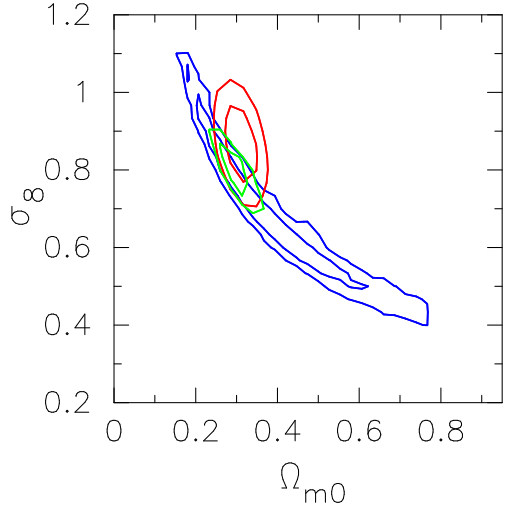}
    \caption{
    	Marginalized 2D confidence level (68\% and 95\%) 
	obtained from cosmic shear data.
	The red region represents the cosmological results by lensing MFs alone 
	and the blue region shows the cosmological constraints by 2PCF alone.
	The green circle shows the result of our combined analysis 
        with the lensing MFs and 2PCF.
	We assume the concordance $\Lambda$CDM model (i.e., $w_0 = -1$) for this figure.
     } 
    \label{fig:LF_MFs+2PCF_LCDM}
    \end{center}
\end{figure}

We would like to examine the ability of the lensing MFs 
to break degeneracies between cosmological parameters.
We first consider the concordance $\Lambda$CDM model, i.e. $w_0 = -1$.
Figure \ref{fig:LF_MFs+2PCF_LCDM} shows the marginalized constraints 
on $\Omega_{\rm 0}$ and $\sigma_{8}$ in the two-parameter plane.
Clearly, the lensing MFs can break the well-known degeneracy 
between $\Omega_{\rm 0}$ and $\sigma_{8}$ that is apparent in the analysis 
using only the 2PCF.
Interestingly, the marginalized constraints on each parameter can be improved 
by a factor of five to eight by adding the MFs.
The final constraints in the case of $\Lambda$CDM model are
summarized in Table \ref{tab:1dcp_lcdm}.

\begin{table}[!t]
\begin{center}
\begin{tabular}{|c|c|c|}
\tableline
& $\Omega_{\rm m0}$ & $\sigma_{8}$  \\ \tableline
2PCF alone & $0.396\pm^{0.177}_{0.185}$ &  $0.695\pm^{0.203}_{0.202}$ \\ \tableline
MFs alone & $0.295\pm0.020$ &  $0.855\pm^{0.060}_{0.060}$ \\ \tableline
MFs + 2PCF & $0.282\pm0.022$ & $0.782\pm^{0.042}_{0.042}$ \\ \tableline
\end{tabular} 
\caption{
  Cosmological parameter constraints obtained from the maximum
  likelihood analysis.
  The error bar shows the 68\% confidence level.
  The concordance $\Lambda$CDM model is assumed for this pot.
  \label{tab:1dcp_lcdm}
}
\end{center}
\end{table}

\begin{figure}[!t]
\begin{center}
      \includegraphics[clip, width=0.45\columnwidth]{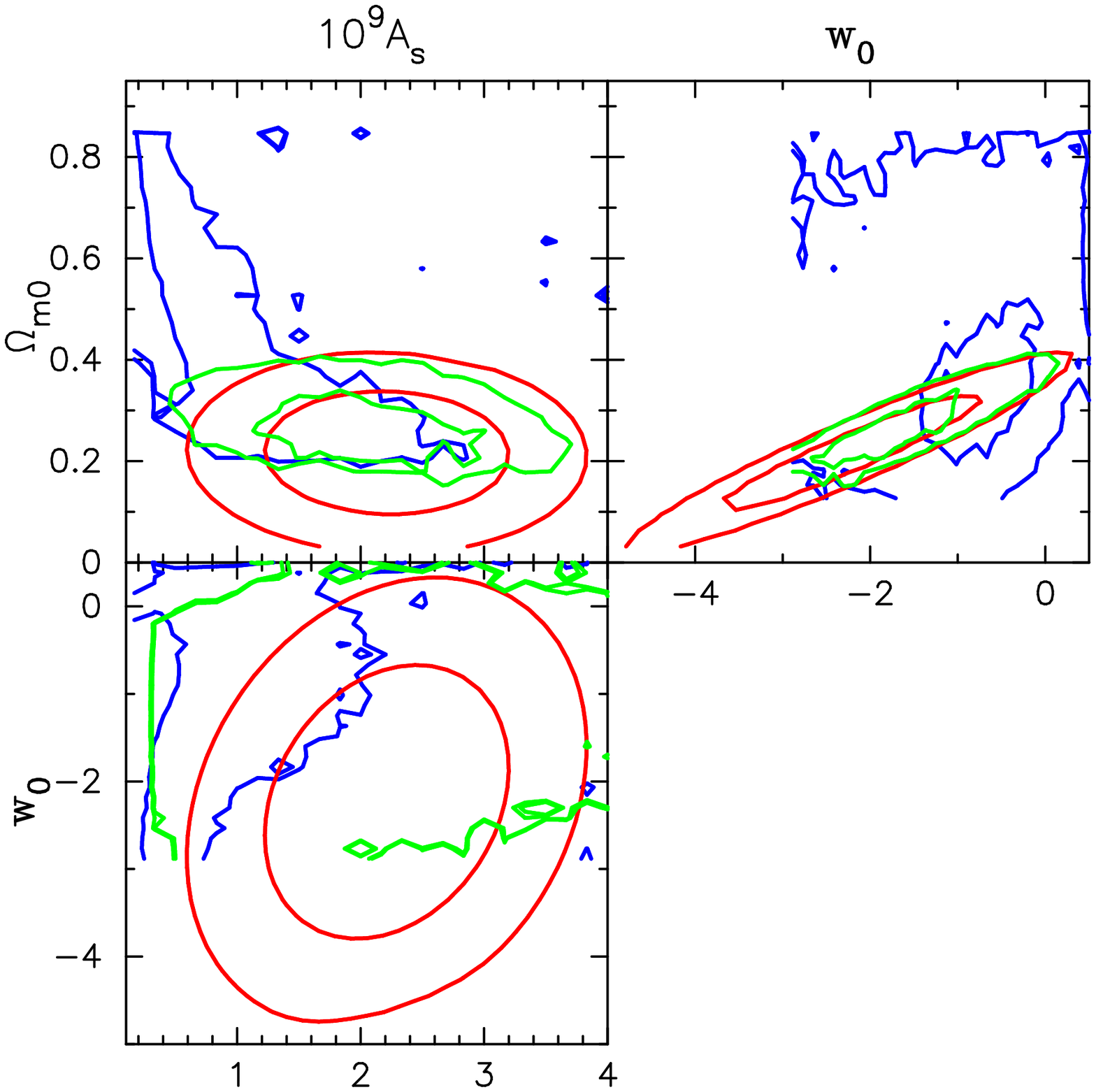}
      \includegraphics[clip, width=0.45\columnwidth]{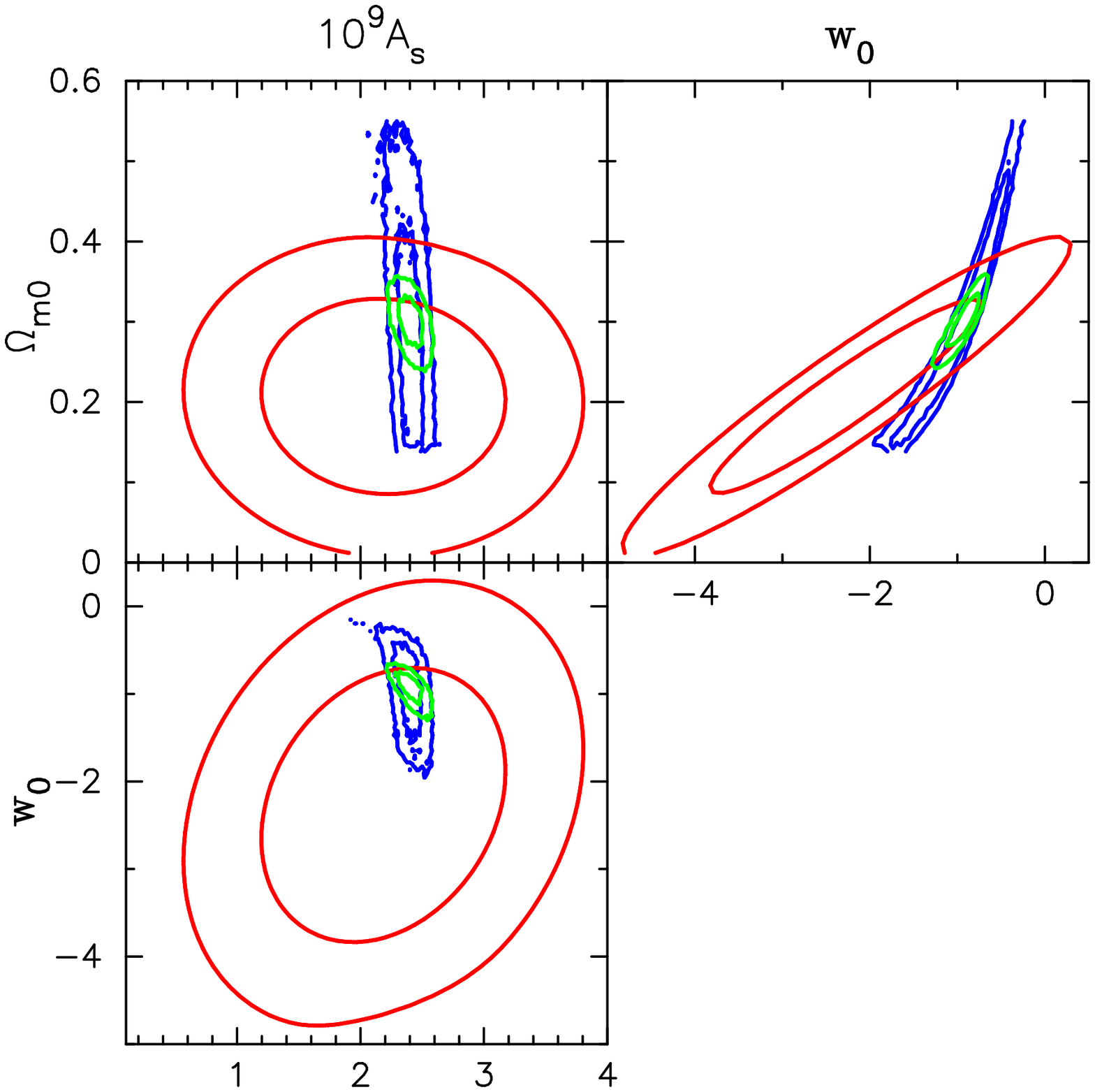}
    \caption{
    	Marginalized 2D confidence level (68\% and 95\%) 
	obtained from the lensing MFs and additional probes.
	The left panel shows the cosmological constraints 
        by our combined analysis with the lensing MFs and
	the 2PCF. The right panel shows the result of 
        the joint analysis with the lensing MFs and CMB.
	In each panel, 
	the red region shows the constraint from the lensing MFs alone 
	and the blue region shows those from the 2PCF or CMB alone.
	The green one shows the result of the combined 
        analysis with the lensing MFs and another data set.	
     } 
    \label{fig:LF_MFs+other}
    \end{center}
\end{figure}

Next, we explore models with a variant of dark energy.
The equation of state parameter $w_0$ serves as an 
additional parameter here.
The left panel in Figure \ref{fig:LF_MFs+other} shows 
the marginalized constraints 
in the 2D plane by the lensing MFs and 2PCF.
The red circle represents the result from the lensing MFs alone,
whereas the green circle
is the estimate derived from combining the lensing MFs and 2PCF.
Interestingly, with the lensing MFs alone, the data set favors 
a low $w_{0}$.
\footnote{
We have also examined which MFs ($V_0, V_1, V_2$) cause this trend.
We have performed likelihood analysis using each MF only.
Both $V_{1}$ and $V_{2}$ prefer lower $w_{0}$.
The 68 $\%$ marginalized constraints on $w_{0}$ using each MF are found to be 
$-0.30\pm^{0.77}_{0.84}$, $-3.31\pm{0.60}$, and $-2.48\pm^{0.91}_{0.84}$ 
for $V_{0}$, $V_{1}$, and $V_{2}$, respectively.
}
We have checked that our theoretical MF template
can recover correctly the input cosmological parameters
for the 40 mock data 
with a similar confidence level expected from the Fisher analysis.
We thus argue that the trend of favoring low $w_0$ 
is likely attributed to the possible systematics as 
studied in Section~\ref{subsec:sys},
or to imperfect modeling of the dependence of the lensing MFs on
the cosmological parameters.

The right panel of Figure \ref{fig:LF_MFs+other} 
shows the 68\% and 95\% confidence regions obtained from
our joint analysis with the WMAP9 CMB data.
The red region represents the results from lensing MFs alone.
The blue region is  
the result obtained from CMB, and the green one represents
constraints by combining the both.
These figures show clearly that the lensing MFs are useful 
to improve the cosmological constraints 
by breaking the parameter degeneracies
The marginalized constraints for the three parameters are 
summarized in Table \ref{tab:1dcp}.
It is worth making further effort to
reduce the systematic errors in measurement of the lensing MFs.

\begin{table}[!t]
\begin{center}
\begin{tabular}{|c|c|c|c|}
\tableline
& $\Omega_{\rm m0}$ & $A_{s}\times 10^9$ & $w_{0}$ \\ \tableline
MFs alone & $0.205\pm0.060$ &  $2.18\pm^{0.60}_{0.60}$ & $-2.2\pm0.8$ \\ \tableline
MFs + 2PCF & $0.256\pm^{0.054}_{0.046}$ & $1.92\pm^{0.65}_{0.65}$ & $-1.60\pm^{0.76}_{0.57}$ \\ \tableline
MFs + CMB & $0.290\pm^{0.016}_{0.028}$ & $2.39\pm0.07$ & $-0.90\pm0.11$ \\ \tableline
\end{tabular} 
\caption{
  We summarize the parameter constraints obtained from the maximum
  likelihood analysis.
  The error bar indicates the 68\% confidence level.
  \label{tab:1dcp}
}
\end{center}
\end{table}

\section{SUMMARY AND CONCLUSION}
\label{sec:con}

We have performed mock lensing observations by incorporating 
the three-dimensional distribution of the source galaxies 
and the effect of imperfect shape measurement in the same manner
as in the analysis of the real CFHTLenS data.
We have then used the mock catalogs and performed a Fisher analysis,
which yields realistic forecast for constraining 
$\Omega_{\rm m0}$, $A_s$, and $w_0$ from the lensing MFs.
We have also studied in detail the possible systematics in the lensing MFs 
measurement that are crucial for cosmological studies.
Finally, we have applied the developed method to real cosmic-shear data,
to show that the lensing MFs are powerful probe of cosmology.

Our analysis using the CFHTLenS data suggests 
that the overall statistical error 
is comparable to the sampling variance for the CHFT survey area
and thus the accuracy of 
cosmological parameter constraints is degraded by a factor of $\sim 2$.
Assuming that the statistical error in upcoming wide-field surveys 
is reduced proportionally to the effective survey area,
we find that the lensing MFs can constrain $w_0$ 
with an error of $\Delta w_0 \sim 0.25$ for
HSC survey with a proposed sky coverage 
of $\sim 1400$ ${\rm deg}^2$.

We have clarified the effects of the two major systematics.
Because of the uncertainties in photometric redshifts of the source galaxies, 
the mean redshift of the lens objects is not determined very accurately.  
We have found that an error of $\Delta z = 0.05$ in the mean source redshift 
results in biased dark energy parameter estimation of $\Delta w_0 \sim 0.2$
for CFHTLenS.
Furthermore, the shear calibration correction causes non-negligible errors
that can bias cosmological parameter estimation as large as 
the 1$\sigma$ confidence level for HSC survey.
Clearly, careful studies are needed in order to understand and correct 
the effects on a survey-by-survey basis.

Weak-lensing MFs are potentially powerful statistics for cosmological studies.
They enable us to constrain cosmological models by cosmic-shear observations even 
without any prior from the cosmic microwave background anisotropies or from the galaxy clustering measurement. There still remain, however, important issues 
when measuring the MFs from real data set.
Although the lensing MFs can be used to extract cosmological information
beyond the two-point statistics, the crucial length scales of structure probed 
are where perturbative approaches break down because of 
the non-linear gravitational growth \citep{2002ApJ...571..638T, 2013PhRvD..88l3002P}.
We need accurate theoretical predictions of the lensing MFs beyond perturbation methods
\citep{2003ApJ...584....1M, 2012MNRAS.419..536M}
in order to sample accurately likelihood functions for a wide range of cosmological parameters.
Another important issue is theoretical uncertainties 
associated with baryonic effects.
Previous studies
\citep[e.g.,][]{2013MNRAS.434..148S,2013PhRvD..87d3509Z}
explored the effect of including baryonic components 
to the 2PCFs and consequently to cosmological parameter estimation.
The baryonic effect could also be important for the MFs analysis
because the MFs generally contain morphological information at arcminute scales, 
i.e., the typical virial radius of galaxy clusters.
\citet{2013PhRvD..87b3511Y} show appreciable baryonic 
effects on peak statistics using a simple model applied 
to dark-matter-only simulations.
Obviously the most straightforward way to include 
the baryonic effect would be to perform 
weak-lensing simulations using outputs of hydrodynamic simulations.
We continue studying the MFs along this idea.

There are other possible systematics than those studied in the present paper.
For example, source-lens clustering \citep[e.g.,][]{2002MNRAS.330..365H}
and the intrinsic alignment \citep[e.g.,][]{2004PhRvD..70f3526H}
are likely to compromise cosmological parameter estimation.
The statistical properties and the correlation 
of source galaxies and lensing structures
are still uncertain but could be critical when making lensing mass maps.
A promising approach in theoretical studies 
would be associating the source positions
with their host dark matter halos on the light cone. 
This is along the line of our ongoing study using
a large set of cosmological simulations in combination with
actual observations.

\acknowledgments
The authors are grateful to Masayuki Tanaka for helpful discussion
on photometric redshift. 
The authors thank Zoltan Haiman 
for useful discussions and comments on the manuscript.
M.S. is supported by Research Fellowships of the Japan Society for 
the Promotion of Science (JSPS) for Young Scientists.
N.Y. acknowledges financial support from
the Japan Society for the Promotion of Science (JSPS) 
Grant-in-Aid for Scientific Research (25287050). 
Numerical computations presented in this paper were in part carried out
on the general-purpose PC farm at Center for Computational Astrophysics,
CfCA, of National Astronomical Observatory of Japan.
This work is based on observations obtained with MegaPrime/MegaCam, 
a joint project of CFHT and CEA/IRFU, at the Canada-France-Hawaii Telescope 
(CFHT) which is operated by the National Research Council (NRC) of Canada, 
the Institut National des Sciences de l'Univers of the Centre National 
de la Recherche Scientifique (CNRS) of France, and the University of Hawaii. 
The research used the facilities of the Canadian Astronomy Data Centre 
operated by the National Research Council of Canada with the support of 
the Canadian Space Agency. CFHTLenS data processing was made possible 
thanks to significant computing support from the NSERC Research Tools 
and Instruments grant program.
\clearpage  

\bibliography{ref}

\begin{thebibliography}{57}
\expandafter\ifx\csname natexlab\endcsname\relax\def\natexlab#1{#1}\fi

\bibitem[{Bartelmann \& Schneider(2001)}]{Bartelmann:1999yn}
Bartelmann, M., \& Schneider, P. 2001, Phys.Rept., 340, 291

\bibitem[{{Ben{\'{\i}}tez}(2000)}]{2000ApJ...536..571B}
{Ben{\'{\i}}tez}, N. 2000, \apj, 536, 571

\bibitem[{{Benjamin} {et~al.}(2013){Benjamin}, {Van Waerbeke}, {Heymans},
  {Kilbinger}, {Erben}, {Hildebrandt}, {Hoekstra}, {Kitching}, {Mellier},
  {Miller}, {Rowe}, {Schrabback}, {Simpson}, {Coupon}, {Fu},
  {Harnois-D{\'e}raps}, {Hudson}, {Kuijken}, {Semboloni}, {Vafaei}, \&
  {Velander}}]{2013MNRAS.431.1547B}
{Benjamin}, J., {Van Waerbeke}, L., {Heymans}, C., {et~al.} 2013, \mnras, 431,
  1547

\bibitem[{{Bennett} {et~al.}(2013){Bennett}, {Larson}, {Weiland}, {Jarosik},
  {Hinshaw}, {Odegard}, {Smith}, {Hill}, {Gold}, {Halpern}, {Komatsu}, {Nolta},
  {Page}, {Spergel}, {Wollack}, {Dunkley}, {Kogut}, {Limon}, {Meyer}, {Tucker},
  \& {Wright}}]{2013ApJS..208...20B}
{Bennett}, C.~L., {Larson}, D., {Weiland}, J.~L., {et~al.} 2013, \apjs, 208, 20

\bibitem[{{Bolzonella} {et~al.}(2000){Bolzonella}, {Miralles}, \&
  {Pell{\'o}}}]{2000A&amp;A...363..476B}
{Bolzonella}, M., {Miralles}, J.-M., \& {Pell{\'o}}, R. 2000, \aap, 363, 476

\bibitem[{{Crocce} {et~al.}(2006){Crocce}, {Pueblas}, \&
  {Scoccimarro}}]{2006MNRAS.373..369C}
{Crocce}, M., {Pueblas}, S., \& {Scoccimarro}, R. 2006, \mnras, 373, 369

\bibitem[{{Eifler} {et~al.}(2009){Eifler}, {Schneider}, \&
  {Hartlap}}]{2009A&A...502..721E}
{Eifler}, T., {Schneider}, P., \& {Hartlap}, J. 2009, \aap, 502, 721

\bibitem[{{Erben} {et~al.}(2001){Erben}, {Van Waerbeke}, {Bertin}, {Mellier},
  \& {Schneider}}]{2001A&A366717E}
{Erben}, T., {Van Waerbeke}, L., {Bertin}, E., {Mellier}, Y., \& {Schneider},
  P. 2001, \aap, 366, 717

\bibitem[{{Erben} {et~al.}(2013){Erben}, {Hildebrandt}, {Miller}, {van
  Waerbeke}, {Heymans}, {Hoekstra}, {Kitching}, {Mellier}, {Benjamin}, {Blake},
  {Bonnett}, {Cordes}, {Coupon}, {Fu}, {Gavazzi}, {Gillis}, {Grocutt}, {Gwyn},
  {Holhjem}, {Hudson}, {Kilbinger}, {Kuijken}, {Milkeraitis}, {Rowe},
  {Schrabback}, {Semboloni}, {Simon}, {Smit}, {Toader}, {Vafaei}, {van Uitert},
  \& {Velander}}]{2013MNRAS.433.2545E}
{Erben}, T., {Hildebrandt}, H., {Miller}, L., {et~al.} 2013, \mnras, 433, 2545

\bibitem[{{Hamana} {et~al.}(2002){Hamana}, {Colombi}, {Thion}, {Devriendt},
  {Mellier}, \& {Bernardeau}}]{2002MNRAS.330..365H}
{Hamana}, T., {Colombi}, S.~T., {Thion}, A., {et~al.} 2002, \mnras, 330, 365

\bibitem[{{Hamana} \& {Mellier}(2001)}]{2001MNRAS.327..169H}
{Hamana}, T., \& {Mellier}, Y. 2001, \mnras, 327, 169

\bibitem[{{Hamana} {et~al.}(2004){Hamana}, {Takada}, \&
  {Yoshida}}]{2004MNRAS.350..893H}
{Hamana}, T., {Takada}, M., \& {Yoshida}, N. 2004, \mnras, 350, 893

\bibitem[{{Hartlap} {et~al.}(2007){Hartlap}, {Simon}, \&
  {Schneider}}]{2007A&A...464..399H}
{Hartlap}, J., {Simon}, P., \& {Schneider}, P. 2007, \aap, 464, 399

\bibitem[{{Heymans} {et~al.}(2012){Heymans}, {Van Waerbeke}, {Miller}, {Erben},
  {Hildebrandt}, {Hoekstra}, {Kitching}, {Mellier}, {Simon}, {Bonnett},
  {Coupon}, {Fu}, {Harnois D{\'e}raps}, {Hudson}, {Kilbinger}, {Kuijken},
  {Rowe}, {Schrabback}, {Semboloni}, {van Uitert}, {Vafaei}, \&
  {Velander}}]{2012MNRAS.427..146H}
{Heymans}, C., {Van Waerbeke}, L., {Miller}, L., {et~al.} 2012, \mnras, 427,
  146

\bibitem[{{Hikage} {et~al.}(2011){Hikage}, {Takada}, {Hamana}, \&
  {Spergel}}]{2011MNRAS.412...65H}
{Hikage}, C., {Takada}, M., {Hamana}, T., \& {Spergel}, D. 2011, \mnras, 412,
  65

\bibitem[{{Hildebrandt} {et~al.}(2012){Hildebrandt}, {Erben}, {Kuijken}, {van
  Waerbeke}, {Heymans}, {Coupon}, {Benjamin}, {Bonnett}, {Fu}, {Hoekstra},
  {Kitching}, {Mellier}, {Miller}, {Velander}, {Hudson}, {Rowe}, {Schrabback},
  {Semboloni}, \& {Ben{\'{\i}}tez}}]{2012MNRAS.421.2355H}
{Hildebrandt}, H., {Erben}, T., {Kuijken}, K., {et~al.} 2012, \mnras, 421, 2355

\bibitem[{{Hinshaw} {et~al.}(2013){Hinshaw}, {Larson}, {Komatsu}, {Spergel},
  {Bennett}, {Dunkley}, {Nolta}, {Halpern}, {Hill}, {Odegard}, {Page}, {Smith},
  {Weiland}, {Gold}, {Jarosik}, {Kogut}, {Limon}, {Meyer}, {Tucker}, {Wollack},
  \& {Wright}}]{2013ApJS..208...19H}
{Hinshaw}, G., {Larson}, D., {Komatsu}, E., {et~al.} 2013, \apjs, 208, 19

\bibitem[{{Hirata} \& {Seljak}(2003)}]{2003MNRAS.343..459H}
{Hirata}, C., \& {Seljak}, U. 2003, \mnras, 343, 459

\bibitem[{{Hirata} \& {Seljak}(2004)}]{2004PhRvD..70f3526H}
{Hirata}, C.~M., \& {Seljak}, U. 2004, \prd, 70, 063526

\bibitem[{{Huterer} {et~al.}(2006){Huterer}, {Takada}, {Bernstein}, \&
  {Jain}}]{2006MNRAS.366..101H}
{Huterer}, D., {Takada}, M., {Bernstein}, G., \& {Jain}, B. 2006, \mnras, 366,
  101

\bibitem[{Kaiser(1992)}]{Kaiser:1991qi}
Kaiser, N. 1992, Astrophys.J., 388, 272

\bibitem[{{Kaiser}(2000)}]{2000ApJ537555K}
{Kaiser}, N. 2000, \apj, 537, 555

\bibitem[{{Kilbinger} {et~al.}(2011){Kilbinger}, {Benabed}, {Cappe}, {Cardoso},
  {Coupon}, {Fort}, {McCracken}, {Prunet}, {Robert}, \&
  {Wraith}}]{2011arXiv1101.0950K}
{Kilbinger}, M., {Benabed}, K., {Cappe}, O., {et~al.} 2011, ArXiv e-prints

\bibitem[{{Kilbinger} {et~al.}(2013){Kilbinger}, {Fu}, {Heymans}, {Simpson},
  {Benjamin}, {Erben}, {Harnois-D{\'e}raps}, {Hoekstra}, {Hildebrandt},
  {Kitching}, {Mellier}, {Miller}, {Van Waerbeke}, {Benabed}, {Bonnett},
  {Coupon}, {Hudson}, {Kuijken}, {Rowe}, {Schrabback}, {Semboloni}, {Vafaei},
  \& {Velander}}]{2013MNRAS.430.2200K}
{Kilbinger}, M., {Fu}, L., {Heymans}, C., {et~al.} 2013, \mnras, 430, 2200

\bibitem[{{Komatsu} {et~al.}(2011){Komatsu}, {Smith}, {Dunkley}, {Bennett},
  {Gold}, {Hinshaw}, {Jarosik}, {Larson}, {Nolta}, {Page}, {Spergel},
  {Halpern}, {Hill}, {Kogut}, {Limon}, {Meyer}, {Odegard}, {Tucker}, {Weiland},
  {Wollack}, \& {Wright}}]{2011ApJS..192...18K}
{Komatsu}, E., {Smith}, K.~M., {Dunkley}, J., {et~al.} 2011, \apjs, 192, 18

\bibitem[{{Kratochvil} {et~al.}(2012){Kratochvil}, {Lim}, {Wang}, {Haiman},
  {May}, \& {Huffenberger}}]{2012PhRvD..85j3513K}
{Kratochvil}, J.~M., {Lim}, E.~A., {Wang}, S., {et~al.} 2012, \prd, 85, 103513

\bibitem[{Lewis {et~al.}(2000)Lewis, Challinor, \& Lasenby}]{Lewis:1999bs}
Lewis, A., Challinor, A., \& Lasenby, A. 2000, Astrophys. J., 538, 473

\bibitem[{{Lim} \& {Simon}(2012)}]{2012JCAP...01..048L}
{Lim}, E.~A., \& {Simon}, D. 2012, \jcap, 1, 48

\bibitem[{Limber(1954)}]{Limber:1954zz}
Limber, D.~N. 1954, Astrophys.J., 119, 655

\bibitem[{{Mandelbaum} {et~al.}(2013){Mandelbaum}, {Slosar}, {Baldauf},
  {Seljak}, {Hirata}, {Nakajima}, {Reyes}, \& {Smith}}]{2013MNRAS.432.1544M}
{Mandelbaum}, R., {Slosar}, A., {Baldauf}, T., {et~al.} 2013, \mnras, 432, 1544

\bibitem[{{Matsubara}(2003)}]{2003ApJ...584....1M}
{Matsubara}, T. 2003, \apj, 584, 1

\bibitem[{{Matsubara} \& {Jain}(2001)}]{2001ApJ...552L..89M}
{Matsubara}, T., \& {Jain}, B. 2001, \apjl, 552, L89

\bibitem[{{Miller} {et~al.}(2013){Miller}, {Heymans}, {Kitching}, {van
  Waerbeke}, {Erben}, {Hildebrandt}, {Hoekstra}, {Mellier}, {Rowe}, {Coupon},
  {Dietrich}, {Fu}, {Harnois-D{\'e}raps}, {Hudson}, {Kilbinger}, {Kuijken},
  {Schrabback}, {Semboloni}, {Vafaei}, \& {Velander}}]{2013MNRAS.429.2858M}
{Miller}, L., {Heymans}, C., {Kitching}, T.~D., {et~al.} 2013, \mnras, 429,
  2858

\bibitem[{Munshi {et~al.}(2008)Munshi, Valageas, Van~Waerbeke, \&
  Heavens}]{Munshi:2006fn}
Munshi, D., Valageas, P., Van~Waerbeke, L., \& Heavens, A. 2008, Phys.Rept.,
  462, 67

\bibitem[{{Munshi} {et~al.}(2012){Munshi}, {van Waerbeke}, {Smidt}, \&
  {Coles}}]{2012MNRAS.419..536M}
{Munshi}, D., {van Waerbeke}, L., {Smidt}, J., \& {Coles}, P. 2012, \mnras,
  419, 536

\bibitem[{{Nishimichi} {et~al.}(2009){Nishimichi}, {Shirata}, {Taruya},
  {Yahata}, {Saito}, {Suto}, {Takahashi}, {Yoshida}, {Matsubara}, {Sugiyama},
  {Kayo}, {Jing}, \& {Yoshikawa}}]{2009PASJ...61..321N}
{Nishimichi}, T., {Shirata}, A., {Taruya}, A., {et~al.} 2009, \pasj, 61, 321

\bibitem[{{Petri} {et~al.}(2013){Petri}, {Haiman}, {Hui}, {May}, \&
  {Kratochvil}}]{2013PhRvD..88l3002P}
{Petri}, A., {Haiman}, Z., {Hui}, L., {May}, M., \& {Kratochvil}, J.~M. 2013,
  \prd, 88, 123002

\bibitem[{{Reid} {et~al.}(2010){Reid}, {Percival}, {Eisenstein}, {Verde},
  {Spergel}, {Skibba}, {Bahcall}, {Budavari}, {Frieman}, {Fukugita}, {Gott},
  {Gunn}, {Ivezi{\'c}}, {Knapp}, {Kron}, {Lupton}, {McKay}, {Meiksin},
  {Nichol}, {Pope}, {Schlegel}, {Schneider}, {Stoughton}, {Strauss}, {Szalay},
  {Tegmark}, {Vogeley}, {Weinberg}, {York}, \& {Zehavi}}]{2010MNRAS.404...60R}
{Reid}, B.~A., {Percival}, W.~J., {Eisenstein}, D.~J., {et~al.} 2010, \mnras,
  404, 60

\bibitem[{{Sato} {et~al.}(2001){Sato}, {Takada}, {Jing}, \&
  {Futamase}}]{2001ApJ...551L...5S}
{Sato}, J., {Takada}, M., {Jing}, Y.~P., \& {Futamase}, T. 2001, \apjl, 551, L5

\bibitem[{{Sato} {et~al.}(2009){Sato}, {Hamana}, {Takahashi}, {Takada},
  {Yoshida}, {Matsubara}, \& {Sugiyama}}]{2009ApJ...701..945S}
{Sato}, M., {Hamana}, T., {Takahashi}, R., {et~al.} 2009, \apj, 701, 945

\bibitem[{{Sato} {et~al.}(2010){Sato}, {Ichiki}, \&
  {Takeuchi}}]{2010PhRvL.105y1301S}
{Sato}, M., {Ichiki}, K., \& {Takeuchi}, T.~T. 2010, Physical Review Letters,
  105, 251301

\bibitem[{{Schneider} {et~al.}(2002){Schneider}, {van Waerbeke}, {Kilbinger},
  \& {Mellier}}]{2002A&A...396....1S}
{Schneider}, P., {van Waerbeke}, L., {Kilbinger}, M., \& {Mellier}, Y. 2002,
  \aap, 396, 1

\bibitem[{{Seitz} \& {Schneider}(1997)}]{1997A&A...318..687S}
{Seitz}, C., \& {Schneider}, P. 1997, \aap, 318, 687

\bibitem[{{Semboloni} {et~al.}(2013){Semboloni}, {Hoekstra}, \&
  {Schaye}}]{2013MNRAS.434..148S}
{Semboloni}, E., {Hoekstra}, H., \& {Schaye}, J. 2013, \mnras, 434, 148

\bibitem[{{Shirasaki} {et~al.}(2013){Shirasaki}, {Yoshida}, \&
  {Hamana}}]{2013ApJ...774..111S}
{Shirasaki}, M., {Yoshida}, N., \& {Hamana}, T. 2013, \apj, 774, 111

\bibitem[{{Shirasaki} {et~al.}(2012){Shirasaki}, {Yoshida}, {Hamana}, \&
  {Nishimichi}}]{2012ApJ...760...45S}
{Shirasaki}, M., {Yoshida}, N., {Hamana}, T., \& {Nishimichi}, T. 2012, \apj,
  760, 45

\bibitem[{{Spergel} {et~al.}(2007){Spergel}, {Bean}, {Dor{\'e}}, {Nolta},
  {Bennett}, {Dunkley}, {Hinshaw}, {Jarosik}, {Komatsu}, {Page}, {Peiris},
  {Verde}, {Halpern}, {Hill}, {Kogut}, {Limon}, {Meyer}, {Odegard}, {Tucker},
  {Weiland}, {Wollack}, \& {Wright}}]{2007ApJS..170..377S}
{Spergel}, D.~N., {Bean}, R., {Dor{\'e}}, O., {et~al.} 2007, \apjs, 170, 377

\bibitem[{{Springel}(2005)}]{2005MNRAS.364.1105S}
{Springel}, V. 2005, \mnras, 364, 1105

\bibitem[{{Takahashi} {et~al.}(2012){Takahashi}, {Sato}, {Nishimichi},
  {Taruya}, \& {Oguri}}]{2012ApJ...761..152T}
{Takahashi}, R., {Sato}, M., {Nishimichi}, T., {Taruya}, A., \& {Oguri}, M.
  2012, \apj, 761, 152

\bibitem[{{Taruya} {et~al.}(2002){Taruya}, {Takada}, {Hamana}, {Kayo}, \&
  {Futamase}}]{2002ApJ...571..638T}
{Taruya}, A., {Takada}, M., {Hamana}, T., {Kayo}, I., \& {Futamase}, T. 2002,
  \apj, 571, 638

\bibitem[{{Tegmark} {et~al.}(2006){Tegmark}, {Eisenstein}, {Strauss},
  {Weinberg}, {Blanton}, {Frieman}, {Fukugita}, {Gunn}, {Hamilton}, {Knapp},
  {Nichol}, {Ostriker}, {Padmanabhan}, {Percival}, {Schlegel}, {Schneider},
  {Scoccimarro}, {Seljak}, {Seo}, {Swanson}, {Szalay}, {Vogeley}, {Yoo},
  {Zehavi}, {Abazajian}, {Anderson}, {Annis}, {Bahcall}, {Bassett}, {Berlind},
  {Brinkmann}, {Budavari}, {Castander}, {Connolly}, {Csabai}, {Doi},
  {Finkbeiner}, {Gillespie}, {Glazebrook}, {Hennessy}, {Hogg}, {Ivezi{\'c}},
  {Jain}, {Johnston}, {Kent}, {Lamb}, {Lee}, {Lin}, {Loveday}, {Lupton},
  {Munn}, {Pan}, {Park}, {Peoples}, {Pier}, {Pope}, {Richmond}, {Rockosi},
  {Scranton}, {Sheth}, {Stebbins}, {Stoughton}, {Szapudi}, {Tucker}, {vanden
  Berk}, {Yanny}, \& {York}}]{2006PhRvD..74l3507T}
{Tegmark}, M., {Eisenstein}, D.~J., {Strauss}, M.~A., {et~al.} 2006, \prd, 74,
  123507

\bibitem[{{Valageas} \& {Nishimichi}(2011)}]{2011A&A...527A..87V}
{Valageas}, P., \& {Nishimichi}, T. 2011, \aap, 527, A87

\bibitem[{{Van Waerbeke} {et~al.}(2013){Van Waerbeke}, {Benjamin}, {Erben},
  {Heymans}, {Hildebrandt}, {Hoekstra}, {Kitching}, {Mellier}, {Miller},
  {Coupon}, {Harnois-D{\'e}raps}, {Fu}, {Hudson}, {Kilbinger}, {Kuijken},
  {Rowe}, {Schrabback}, {Semboloni}, {Vafaei}, {van Uitert}, \&
  {Velander}}]{2013MNRAS.433.3373V}
{Van Waerbeke}, L., {Benjamin}, J., {Erben}, T., {et~al.} 2013, \mnras, 433,
  3373

\bibitem[{{White} \& {Hu}(2000)}]{2000ApJ...537....1W}
{White}, M., \& {Hu}, W. 2000, \apj, 537, 1

\bibitem[{{Wraith} {et~al.}(2009){Wraith}, {Kilbinger}, {Benabed}, {Capp{\'e}},
  {Cardoso}, {Fort}, {Prunet}, \& {Robert}}]{2009PhRvD..80b3507W}
{Wraith}, D., {Kilbinger}, M., {Benabed}, K., {et~al.} 2009, \prd, 80, 023507

\bibitem[{{Yang} {et~al.}(2013){Yang}, {Kratochvil}, {Huffenberger}, {Haiman},
  \& {May}}]{2013PhRvD..87b3511Y}
{Yang}, X., {Kratochvil}, J.~M., {Huffenberger}, K., {Haiman}, Z., \& {May}, M.
  2013, \prd, 87, 023511

\bibitem[{{Zentner} {et~al.}(2013){Zentner}, {Semboloni}, {Dodelson}, {Eifler},
  {Krause}, \& {Hearin}}]{2013PhRvD..87d3509Z}
{Zentner}, A.~R., {Semboloni}, E., {Dodelson}, S., {et~al.} 2013, \prd, 87,
  043509

\end{thebibliography}

\clearpage 

\appendix
\section{EFFECT OF SOURCE REDSHIFT CLUSTERING 
ON THE VARIANCE OF SMOOTHED CONVERGENCE FIELD}

Here, we summarize the effect of source redshift clustering on 
the variance of a smoothed convergence field.
Weak-lensing convergence $\kappa$ is given 
by the integral of matter over density with a weight
along the line-of-sight $\bd{\theta}$:
\beqa
\kappa(\bd{\theta},\chi_s)&=&
\frac{3}{2}\left(\frac{H_{0}}{c}\right)^2 \Omega_{\rm m0}
\int _{0}^{\chi_s}{\rm d}\chi_{\ell} \ g(\chi_{s}, \chi_{\ell})
\frac{\delta[r(\chi_{\ell})\bd{\theta},\chi_{\ell}]}{a(\chi_{\ell})}, \label{eq:kappa_delta_a} \\
g(\chi_s, \chi_{\ell}) &=&
\frac{r(\chi_s - \chi_{\ell})r(\chi_{\ell})}{r(\chi_s)},
\eeqa
where $\chi$ is comoving distance, $r(\chi)$ is angular diameter distance, 
and $\chi_{s}$ represents the comoving distance to a source.
One can assign a probability distribution $p(\chi_s)$  
of a source galaxy's position, or in fact $p(\chi_s)$ 
for a population of source galaxies,  
and integrate as,
\beqa
\bar{\kappa}(\bd{\theta})= \frac{3}{2}\left(\frac{H_{0}}{c}\right)^2 \Omega_{\rm m0}
\int_{0}^{\chi_{H}}{\rm d}\chi_s \ p(\chi_s) \int _{0}^{\chi_s}{\rm d}\chi_{\ell} \ g(\chi_{s}, \chi_{\ell})
\frac{\delta[r(\chi_{\ell})\bd{\theta},\chi_{\ell}]}{a(\chi_{\ell})}. \label{eq:bar_kappa_delta}
\eeqa
In the conventional multiple lens plane algorithm 
\citep{2000ApJ...537....1W, 2001MNRAS.327..169H},
one can calculate the both convergence field $\kappa$ and $\bar{\kappa}$ 
by using a suitable weight function in the integral.
In practice in ray-tracing simulations, we shoot rays from
the observer point to the source redshifts to obtain $\kappa$,
whereas we shoot rays up to some certain (high-)redshift
but with weight $p(\chi_s)$ along the line-of-sights to obtain 
$\bar{\kappa}$.
In the former case, the full three-dimensional positions
of the source galaxies are realized as in the
observation considered, i.e., CFHTLenS in our case. 
The difference between $\kappa$ 
and $\bar{\kappa}$ can be easily seen in a direct manner
using the two sets of simulations,
and the resulting variances of smoothed 
convergence field can be explicitly compared.

The smoothed convergence field for $\kappa$ is given by
\beqa
{\cal K}(\bd{\theta}) = \sum_{i} U(\bd{\theta}-\bd{\phi}_{i})\kappa(\bd{\phi}_{i}, \chi_{si}), \label{eq:sm_kap}
\eeqa
where $U(\bd{\theta})$ is the filter function for smoothing and 
the summation is taken over the source objects.
The smoothed convergence field for $\bar{\kappa}$ is also 
obtained in the same way.
The two-point correlation function of $\cal{K}$ is then given by
\beqa
\langle {\cal K}(\bd{\theta}_{1}) {\cal K}(\bd{\theta}_{2}) \rangle
&=&\langle \sum_{i} \sum_{j}U(\bd{\theta}_{1}-\bd{\phi}_{i})U(\bd{\theta}_{2}-\bd{\phi}_{j})
\kappa(\bd{\phi}_{i},\chi_{si}) \kappa(\bd{\phi}_{j},\chi_{sj}) \rangle \nonumber \\ 
&=&\int {\rm d}^2 \phi_{1}{\rm d}^2 \phi_{2}\,U(\bd{\theta}_{1}-\bd{\phi}_{1})U(\bd{\theta}_{2}-\bd{\phi}_{2}) \nonumber \\
&\times&\int {\rm d}\chi_{s1} \,{\rm d}\chi_{s2} \ p(\chi_{s1})p(\chi_{s2})
\left[1+\xi_{ss}(\bd{\phi}_{1}-\bd{\phi}_{2},\chi_{s1}, \chi_{s2})\right] \nonumber \\
&&\hspace{150pt}\times\langle \kappa(\bd{\phi}_{1},\chi_{s1})\kappa(\bd{\phi}_{2},\chi_{s2})\rangle, \label{eq:sm_kap_2pcf}
\eeqa
where $\langle \cdots \rangle$ represents the operator of ensemble average and
$\xi_{ss}$ represents the two point correlation function of the sources.
One can also calculate the two-point correlation 
of $\bar{\cal K}$ in the similar manner.
We then obtain
the non-vanishing difference between $\langle \bar{\cal K}\bar{\cal K}\rangle$ 
and $\langle {\cal K}{\cal K}\rangle$
as
\beqa
\langle \bar{\cal K}\bar{\cal K} - {\cal K}{\cal K}\rangle
&=&\int {\rm d}^2 \phi_{1}{\rm d}^2 \phi_{2} \ U(\bd{\theta}_{1}-\bd{\phi}_{1})U(\bd{\theta}_{2}-\bd{\phi}_{2}) \nonumber \\
&\times& \left[{\bar{\xi}_{ss}}(\bd{\phi}_{1}-\bd{\phi}_{2})w_{\rm pp}(\bd{\phi}_{1}-\bd{\phi}_{2})
-V_{\rm pp}(\bd{\phi}_{1}-\bd{\phi}_{2})\right], \label{eq:nonvanish_2pcf} \\
\bar{\xi}_{ss}(\bd{\phi}_{1}-\bd{\phi}_{2})
&=&
\int {\rm d}\chi_{s1} {\rm d}\chi_{s2} \ p(\chi_{s1})p(\chi_{s2})
\xi_{ss}(\bd{\phi}_{1}-\bd{\phi}_{2},\chi_{s1}, \chi_{s2}), \\
w_{\rm pp}(\bd{\phi}_{1}-\bd{\phi}_{2})
&=& \langle \bar{\kappa}(\bd{\phi}_{1})\bar{\kappa}(\bd{\phi}_{2})\rangle, \\
V_{\rm pp}(\bd{\phi}_{1}-\bd{\phi}_{2})
&=&\frac{9}{4}\left(\frac{H_{0}}{c}\right)^4 \Omega_{\rm m0}^2
\int{\rm d}\chi_{s1} {\rm d}\chi_{s2} \ p(\chi_{s1})p(\chi_{s2})
\xi_{ss}(\bd{\phi}_{1}-\bd{\phi}_{2},\chi_{s1}, \chi_{s2}) \nonumber \\
&\times&\int {\rm d}\chi_{\ell 1} {\rm d}\chi_{\ell 2} \ 
\frac{g(\chi_{s1}, \chi_{\ell 1})g(\chi_{s2},\chi_{\ell 2})}{a(\chi_{\ell 1})a(\chi_{\ell 2})}
\langle
\delta[r(\chi_{\ell 1})\bd{\phi_{1}},\chi_{\ell 1}]
\delta[r(\chi_{\ell 2})\bd{\phi_{2}},\chi_{\ell 2}]
\rangle.
\eeqa
This non-vanishing term arises 
if the source clustering $\xi_{ss}$ evolves over redshift.
Note also that the MFs of $\bar{\cal K}$ and those of
${\cal K}$ can be, in general, different if their 
variances differ \citep[see, e.g.][]{2013ApJ...774..111S}. 

In practice, the smoothed convergence field is often 
estimated from the shear field $\gamma$.
In this case, one can calculate ${\cal K}$ using the following equation,
\beqa
{\cal K}(\bd{\theta}) = \sum_{i} Q_{t}(\bd{\theta}-\bd{\phi}_{i})\gamma_{t}(\bd{\phi}_{i}, \chi_{si}), 
\label{eq:sm_kap_v2}
\eeqa
where $Q_{t}(\bd{\theta})$ is the filter function for the shear field
which is related to $U(\bd{\theta})$ 
by Equation~(\ref{eq:U_Q_fil}) and $\gamma_{t}(\bd{\theta}, \chi_{s})$ is the tangential component of shear at the position $\bd{\theta}$ when a source is at 
$\chi_{s}$ from the observer.
Using Equation~(\ref{eq:sm_kap_v2}), one can derive
the correspoding non-vanishing term
\beqa
\langle \bar{\cal K}\bar{\cal K} - {\cal K}{\cal K}\rangle
&=&\int {\rm d}^2 \phi_{1}{\rm d}^2 \phi_{2} \ Q_{t}(\bd{\theta}_{1}-\bd{\phi}_{1})
Q_{t}(\bd{\theta}_{2}-\bd{\phi}_{2}) \nonumber \\
&\times&\left[\bar{\xi}_{ss}(\bd{\phi}_{1}-\bd{\phi}_{2})
\langle \bar{\gamma_{t}}(\bd{\phi}_{1})\bar{\gamma_{t}}(\bd{\phi}_{2}) \rangle
-V_{\rm pp}^{\gamma}(\bd{\phi}_{1}-\bd{\phi}_{2})\right], \label{eq:nonvanish_2pcf_v2} \\
V_{\rm pp}^{\gamma}(\bd{\phi}_{1}-\bd{\phi}_{2})
&=&
\int{\rm d}\chi_{s1} {\rm d}\chi_{s2} \ p(\chi_{s1})p(\chi_{s2})
\xi_{ss}(\bd{\phi}_{1}-\bd{\phi}_{2},\chi_{s1}, \chi_{s2}) \nonumber \\
&\times& 
\langle \gamma_{t}(\bd{\phi}_{1},\chi_{s1})\gamma_{t}(\bd{\phi}_{2}, \chi_{s2}) \rangle, \\
\langle \bar{\gamma_{t}}(\bd{\phi}_{1})\bar{\gamma_{t}}(\bd{\phi}_{2}) \rangle
&=&\frac{9}{4}\left(\frac{H_{0}}{c}\right)^4 \Omega_{\rm m0}^2
\int_{0}^{\chi_{H}}\frac{{\rm d\chi}}{a^2(\chi)}{\bar W}^2(\chi) \nonumber \\
&\times&
\int\frac{\ell d\ell}{2\pi}P_{\delta}\left(k=\frac{\ell}{r(\chi)},z(\chi)\right)
\left(\frac{J_{0}(\ell\phi_{12})+J_{4}(\ell\phi_{12})}{2}\right), \\
{\bar W}(\chi)
&=&\int_{\chi}^{\chi_{H}}{\rm d\chi^{\prime}}
p(\chi^{\prime})\frac{r(\chi^{\prime}-\chi)}{r(\chi^{\prime})}, \\
\langle \gamma_{t}(\bd{\phi}_{1},\chi_{s1})\gamma_{t}(\bd{\phi}_{2}, \chi_{s2}) \rangle
&=&
\frac{9}{4}\left(\frac{H_{0}}{c}\right)^4 \Omega_{\rm m0}^2
\int_{0}^{{\rm min}(\chi_{s1}, \chi_{s2})}\frac{{\rm d\chi}}{a^2(\chi)}
\frac{r(\chi_{s1}-\chi)}{r(\chi_{s1})}\frac{r(\chi_{s2}-\chi)}{r(\chi_{s2})} \nonumber \\
&\times&
\int\frac{\ell d\ell}{2\pi}P_{\delta}\left(k=\frac{\ell}{r(\chi)},z(\chi)\right)
\left(\frac{J_{0}(\ell\phi_{12})+J_{4}(\ell\phi_{12})}{2}\right),
\eeqa
where $\phi_{12}$ is the norm of $\bd{\phi}_{1}-\bd{\phi}_{2}$ and 
$P_{\delta}(k,z)$ is the non-linear matter power spectrum at redshift $z$.

\begin{figure}[!t]
\begin{center}
       \includegraphics[clip, width=0.32\columnwidth]{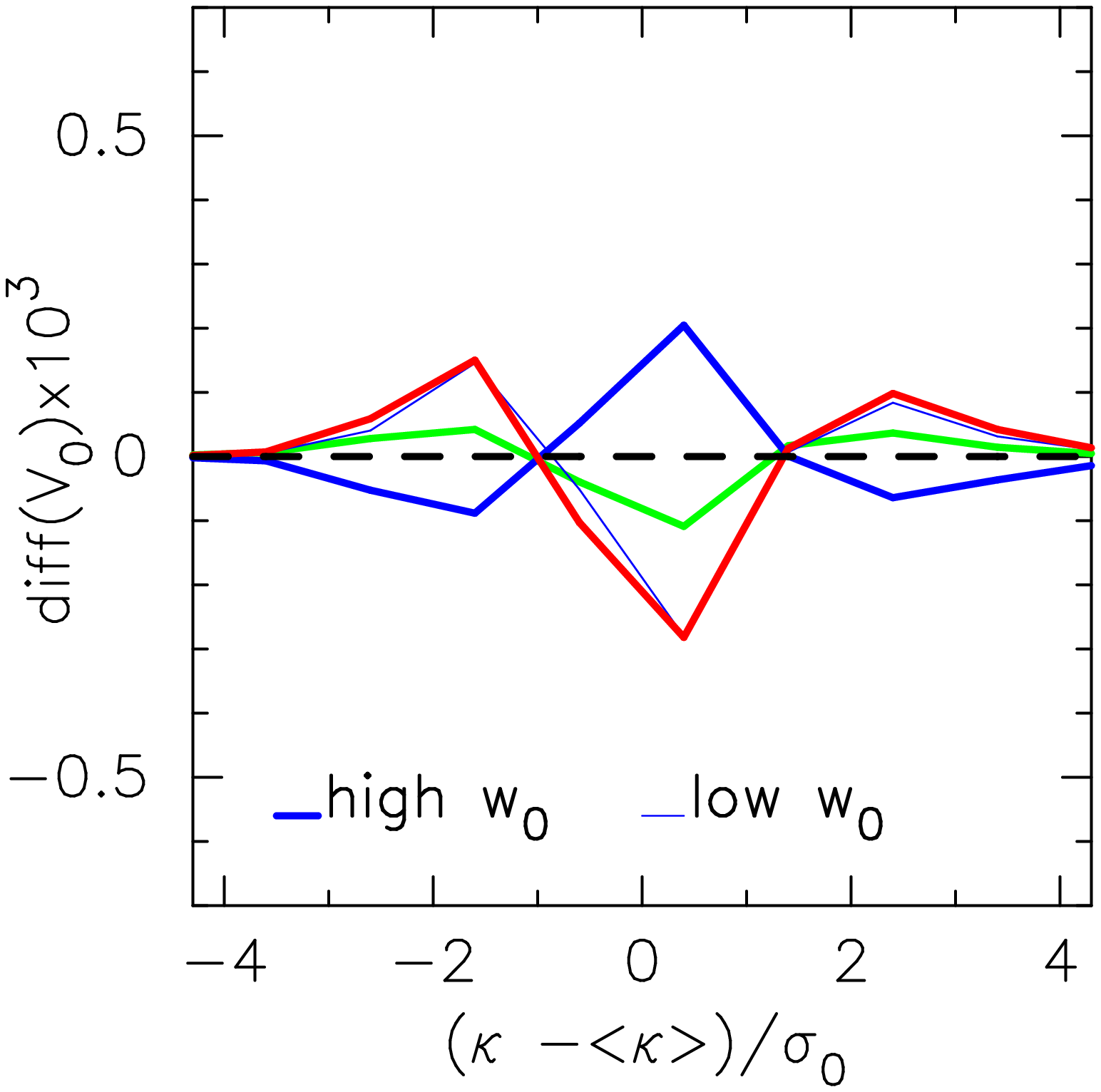}
       \includegraphics[clip, width=0.32\columnwidth]{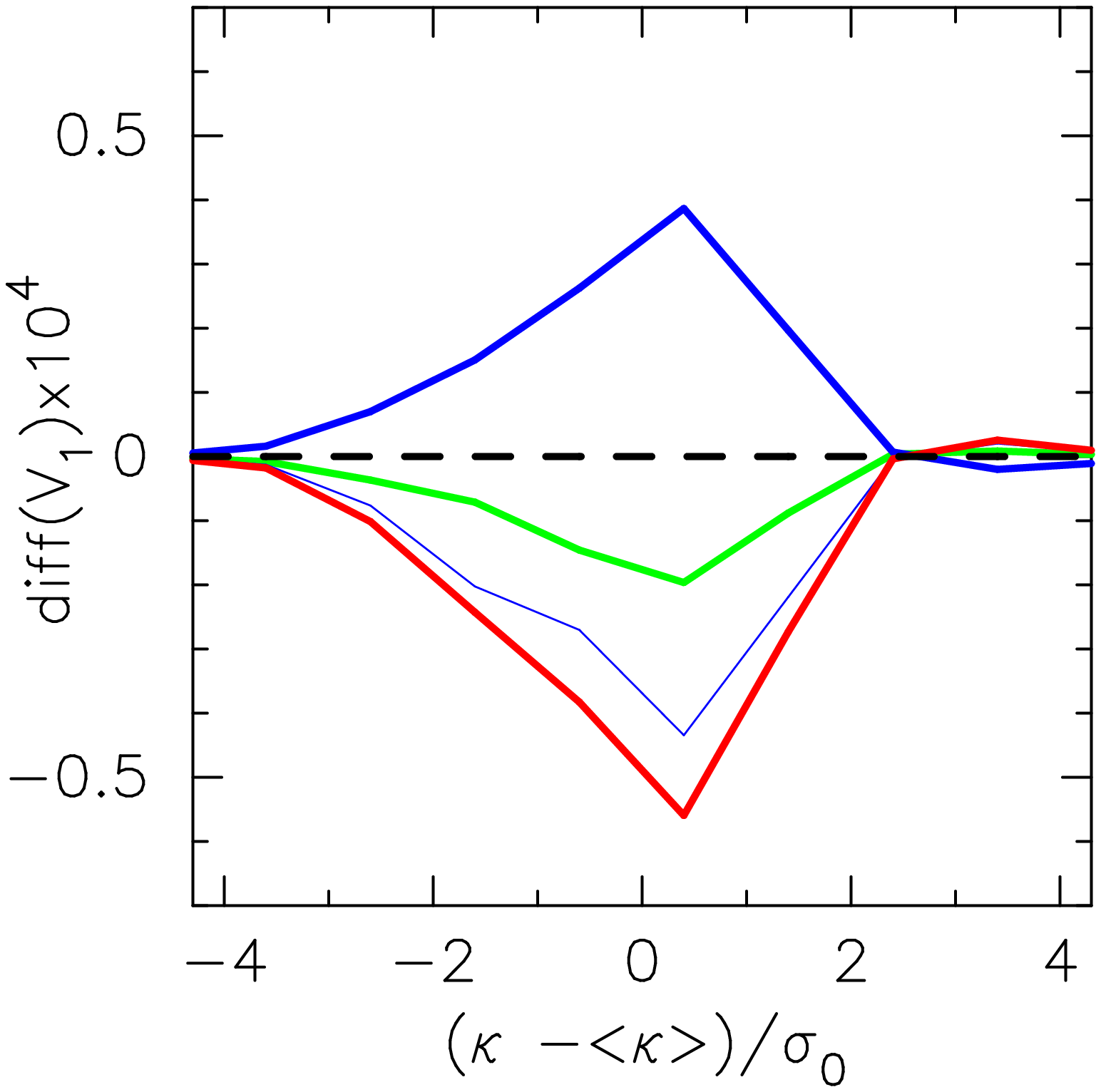}
       \includegraphics[clip, width=0.32\columnwidth]{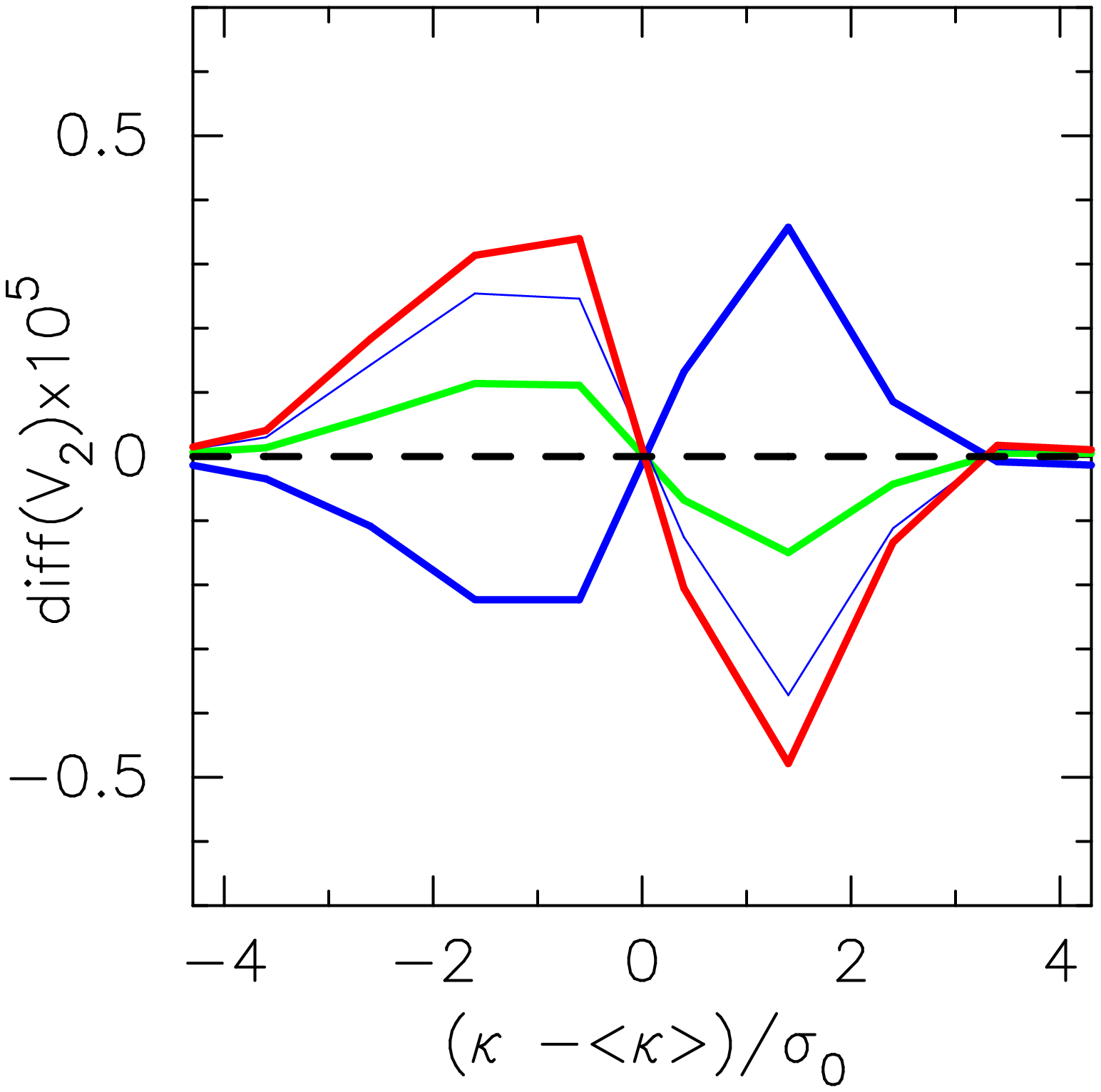}
    \caption{
	Impact of a source redshift clustering on lensing MFs.
	We plot the differences of average MFs over 40 catalogs between our fiducial cosmology and another one 
	that includes a given systematic.
	We generate a new set of mock catalogs in order to include systematics in our analysis.
	The red line shows the impact of the difference of source redshift distribution
	and the green one represents the effect of source redshift clustering on lensing MFs.
	For comparison, the thick (thin) blue line corresponds to the case of cosmological model 
	with higher (lower) $w_0$.
     } 
    \label{fig:diffMFs_sys_zdist}
    \end{center}
\end{figure}

Although we have derived the difference between ${\cal K}$ and ${\bar{\cal K}}$
at the two-point statistics, 
it is extremely difficult to derive an explicit form for the
corresponding difference in the lensing MFs.
We thus resort to comparing directly 
the two sets of simulated lensing MFs.
One is our fiducial mock data used in Section~\ref{subsec:mock}.
For the other, new set of simulations, we calculate $\bar{\kappa}$ 
at each source position on the sky using the source redshift 
distribution (weight) that is shown as the black histograms in Figure \ref{fig:zdist}.
Figure \ref{fig:diffMFs_sys_zdist} shows the results.
The red line shows the difference caused by the two different source redshift 
distributions as described in Section~\ref{subsec:sys}.
The green line represents the difference of the lensing MFs with and 
without source redshift clustering.
For reference, we also plot the difference of lensing MFs 
between the different cosmological model by blue lines.
The thick (thin) blue lines correspond to the case of the 
cosmological model with higher (lower) $w_0$.
Although the impact of source clustering (green) is
smaller than the effect of different source distribution (red),
it or actually both could be a major source of systematics for
future survey with the sky coverage of 20000 ${\rm deg}^2$.
The induced biases in cosmological parameters due to the source clustering
are estimated by Equation~(\ref{eq:bias_CP});
the results are 
$\Delta \Omega_{\rm m0} = 0.00642$,
$\Delta A_{s} = -0.00467 \times 10^9$
and $\Delta w_{0} = 0.00487$.

\section{ESTIMATING THE MINKOWSKI FUNCTIONALS COVARIANCE MATRIX}
We describe an approximate way to evaluate the covariance matrix
as given by Equation~(\ref{eq:Cov}) and test its validity in this Appendix.
We first generate 40 noise-free lensing maps by the method in Section~ \ref{subsec:rt}.
For each realization of the 40 maps, 
we use a different random seed for the intrinsic ellipticities
to make mock source galaxy catalogs described in Section~\ref{subsec:mock}.
In this way, we generate $40\times40=1600$ catalogs in total,
which can be used to estimate the full covariance of the MFs.
\begin{table}[!t]
\begin{center}
\begin{tabular}{|c|c|c|c|c|c|}
\hline
& $x_{1}$ & $x_{2}$ & $x_{3}$ & $x_{4}$ & $x_{5}$\\ \hline
$x_{1}$ & 1.96 &  0.35 & 1.96 & 4.07 & 1.87  \\ \hline
$x_{2}$ &  & 1.53 & 1.51 & 1.27 & 1.46 \\ \hline
$x_{3}$ &  &  & 1.82 & 2.12 & 1.92 \\ \hline
$x_{4}$ &  &  &  & 1.14 & 0.98  \\ \hline
$x_{5}$ &  &  &  &  & 1.82 \\ \hline
\end{tabular} 
\caption{
The ratio of the full covariance of $V_{0}$ to our estimator.
The full covariance is derived from the new 1600 maps with Equation~(\ref{eq:c_full}),
whereas our estimator is given by Equation~(\ref{eq:c_ours}).
}
\label{tab:check_cov}
\end{center}
\end{table}
Let us denote a mock catalogue as ${\cal K}^{m,n}$, 
which is generated by the $m$th noise free lensing map 
with an $n$th random seed of the intrinsic ellipticity distribution.
We then calculate the full covariance of $V_{0}$ as follows:
\beqa
C^{(40, 40)}_{ij}
&\equiv&
\frac{1}{1600-1}\sum_{m=1}^{40}\sum_{n=1}^{40}
\sum_{i,j}(V_{0}(x^{m,n}_{i})-\bar{V}_{0}(x^{m,n}_{i}))(V_{0}(x^{m,n}_{j})-\bar{V}_{0}(x^{m,n}_{j})), \label{eq:c_full} \\ 
\bar{V}_{0}(x^{m,n}_{i}) 
&\equiv& \frac{1}{1600}\sum_{m=1}^{40}\sum_{n=1}^{40}V_{0}(x^{m,n}_{i}),
\eeqa
where $x^{m,n}_{i} = ({\cal K}_{i}^{m,n}-\langle {\cal K}^{m,n}\rangle)/\sigma_{0}^{m,n}$ 
and we here use five bins in the range of $x_{i}=[-3:3]$.
We also calculate our estimator adopted in this paper:
\beqa
C_{ij} &=& C^{(40, 1)}_{ij}+C^{(1, 40)}_{ij}, \label{eq:c_ours}
\eeqa
where
\beqa
C^{(1, 40)}_{ij}
&\equiv&
\frac{1}{40-1}\sum_{n=1}^{40}
\sum_{i,j}
(V_{0}(x^{1,n}_{i})-\bar{V}_{0}(x^{1,n}_{i}))(V_{0}(x^{1,n}_{j})-\bar{V}_{0}(x^{1,n}_{j})), \\
C^{(40, 1)}_{ij}
&\equiv&
\frac{1}{40-1}\sum_{m=1}^{40}
\sum_{i,j}
(V_{0}(x^{m,1}_{i})-\bar{V}_{0}(x^{m,1}_{i}))(V_{0}(x^{m,1}_{j})-\bar{V}_{0}(x^{m,1}_{j})), \\
\bar{V}_{0}(x^{1,n}_{i}) 
&\equiv& \frac{1}{40}\sum_{n=1}^{40}V_{0}(x^{1,n}_{i}),\\
\bar{V}_{0}(x^{m,1}_{i}) 
&\equiv& \frac{1}{40}\sum_{m=1}^{40}V_{0}(x^{m,1}_{i}).
\eeqa

The ratio $C^{(40, 40)}_{ij}/C_{ij}$ then serves a check 
on the accuracy of our estimator.
We summarize the ratio for each component in Table \ref{tab:check_cov}.
Our estimator indeed gives a good approximation to 
the full covariance.
The ratio is typically within a factor of two,
and the same is also found for $V_{1}$ and $V_{2}$.
Even if the ratio of the full covariance and our estimator is 2
for all the matrix elements, 
the cosmological forecast shown in this paper
would be degraded only by a factor of $\sim 2^{1/3}$ (i.e., $\sim20\%$).

\end{document}